\documentclass[preprint,pdftex]{elsarticle}

\usepackage{graphicx}
\usepackage{color}
\usepackage{amsmath,amssymb}
\usepackage{booktabs}
\usepackage{url}
\usepackage{lineno}
\usepackage{hyperref}

\journal{Speech Communication}

\biboptions{authoryear}

\begin{document}

\begin{frontmatter}

\title{GEDI: Gammachirp Envelope Distortion Index for Predicting Intelligibility of Enhanced Speech}


\author[WU]{Katsuhiko Yamamoto}
\ead{yamamoto.katsuhiko@g.wakayama-u.jp}

\author[WU]{Toshio Irino}
\ead{irino@wakayama-u.ac.jp}

\author[NTT]{Shoko Araki}
\ead{araki.shoko@lab.ntt.co.jp}

\author[NTT]{Keisuke Kinoshita}
\ead{kinoshita.k@lab.ntt.co.jp}

\author[NTT]{Tomohiro Nakatani}
\ead{nakatani.tomohiro@lab.ntt.co.jp}

\address[WU]{Graduate School of Systems Engineering, Wakayama University, Sakaedani 930, Wakayama, Wakayama 640--8510, Japan}
\address[NTT]{NTT Communication Science Laboratories, 2-4 Hikaridai, Seika-cho, Soraku-gun, Kyoto 619--0237, Japan}

\begin{abstract}
In this study, we propose a new concept, the gammachirp envelope distortion index (GEDI), based on the signal-to-distortion ratio in the auditory envelope, $\mathrm{SDR_{env}}$, to predict the intelligibility of speech enhanced by nonlinear algorithms. The objective of GEDI is to calculate the distortion between enhanced and clean-speech representations in the domain of a temporal envelope extracted by the gammachirp auditory filterbank and modulation filterbank. We also extend GEDI with multi-resolution analysis (mr-GEDI) to predict the speech intelligibility of sounds under non-stationary noise conditions. We evaluate GEDI in terms of the speech intelligibility predictions of speech sounds enhanced by a classic spectral subtraction and a Wiener filtering method. The predictions are compared with human results for various signal-to-noise ratio conditions with additive pink and babble noises. The results showed that mr-GEDI predicted the intelligibility curves better than short-time objective intelligibility (STOI) measure, extended-STOI (ESTOI) measure, and hearing-aid speech perception index (HASPI) under pink-noise conditions, and better than HASPI under babble-noise conditions. The mr-GEDI method does not present an overestimation tendency and is considered a more conservative approach than STOI and ESTOI. Therefore, the evaluation with mr-GEDI may provide additional information in the development of speech enhancement algorithms.
\end{abstract}


\end{frontmatter}


\section{Introduction}
\label{sec:Intro}
The development of objective speech intelligibility and quality measures is essential for speech communication technologies, such as assistive listening devices (e.g., smart headphones and hearing aids) \citep{Falk2015}. International standards of objective intelligibility measures (OIM), the speech intelligibility index (SII) \citep{ANSI_S3-5_1997}, and the speech transmission index (STI) \citep{ISO_9921_2003} have been proposed to evaluate speech transmission qualities of public spaces and telecommunication lines assumed to be linear transmission systems. However, SII and STI cannot account for the effects of nonlinear processing, including noise reduction and speech-enhancement algorithms. For example, it was reported that STI failed to predict speech intelligibility of enhanced speech processed by a simple spectral subtraction (SS) algorithm \citep{Jorgensen2011}. For the above reasons, the evaluation methodology of speech intelligibility still involves subjective listening tests. However, many noise reduction and speech-enhancement algorithms have been developed thus far without properly designed subjective listening tests evaluation. 

\subsection{Objective intelligibility measures for speech enhancement}
\label{sec:Intro_OIM}

To solve these problems, several human auditory-based models have been proposed. 
\cite{Rhebergen2005} extended SII using short-time calculations of signal-to-noise ratios (SNRs) in the spectral domain.
\cite{Kates2005} replaced the SNR with a signal-to-distortion ratio (SDR), calculated from the magnitude-squared coherence function using cross-spectrum between clean speech ($S$) and enhanced speech ($\hat{S}$). This method was referred to as the coherence speech intelligibility index (CSII). \cite{Cooke2006} proposed the ``glimpsing model'' in human auditory models, which calculates local SNRs of spectro-temporal excitation patterns in the time-frequency domain. 

Whereas the above models calculated indices in the spectral domain, there were alternative OIMs based on the calculation in the envelope modulation domain. 
They also used correlation analysis or SNR.
\cite{Taal2011} proposed a short-time objective intelligibility (STOI) measure, which has often been used in evaluations of speech enhancement algorithms. STOI is based on the cross-correlation between the temporal envelopes of clean speech ($S$) and enhanced speech ($\hat{S}$) at the output of a 1/3-octave filterbank. 
STOI assesses the intelligibility of speech processed by ideal time-frequency segregation (ITFS) \citep{Kjems2009} and minimum mean-squared error estimate of short-time spectral amplitude algorithms \citep{Ephraim1985,Erkelens2007}. 
STOI was evaluated under speech-shaped noise (SSN), cafeteria noise, noise from a bottling factory hall and noise conditions of the interior of a car \citep{Taal2011}.
\cite{Jensen2016} proposed an extended version of STOI to account for highly modulated masker conditions, where STOI could not capture the spectral and temporal correlations of speech from the noise. 
Extended STOI (ESTOI) calculates the index from the spectral correlation of sub-band envelopes, whereas the original STOI calculated it directly from temporal correlations.
It was reported that ESTOI demonstrated a performance comparable to STOI and also showed an overall improved prediction performance under the conditions of temporally modulated maskers \citep{Dreschler2001,Gustafsson1994} and recorded noise in the Noisex corpus \citep{Varga1993} and ICRA noises \citep{Dreschler2001}, where STOI did not perform well.

\cite{Kates2014} proposed a hearing-aid speech perception index (HASPI) for hearing-impaired (HI) and normal-hearing (NH) listeners as an extension of CSII \citep{Kates2005}. This measure combined two indices: the coherence between the outputs of an auditory filterbank for clean ($S$) with enhanced speech ($\hat{S}$) and the cross-correlation between the temporal sequences of the cepstral coefficients of $S$ and $\hat{S}$.  
HASPI was said to account for nonlinear frequency compression, ITFS processing, and noise vocoded speech under multi-talker babble noise and SSN conditions.

\cite{Jorgensen2011} proposed an alternative SNR-based model, referred to as the speech-based envelope power-spectrum model (sEPSM). sEPSM assumes that speech intelligibility is related to the SNR in the envelope domain, ${\rm SNR_{env}}$, which originates from ${\rm (S/N)_{mod}}$, as shown in \citep{Dubbelboer2008}. ${\rm SNR_{env}}$ is calculated from the ratios between the envelope powers of the enhanced speech ($\hat{S}$) and the residual noise ($\tilde{N}$) in the modulation frequency domain. sEPSM is intended to assess the intelligibility of speech sounds processed by SS. sEPSM was extended to a multi-resolution version to perform more accurate speech intelligibility estimations for speech affected by non-stationary noises \citep{Jorgensen2013}. \cite{Chabot-Leclerc2014} extended sEPSM with a spectro-temporal receptive field to account for phase jitter \citep{Chi1999}. 
sEPSM process is useful for developing a non-intrusive OIM, because sEPSM calculates the index without using an original clean signal.
\cite{Santos2014} proposed a non-intrusive OIM to predict speech intelligibility in reverberant environments using the gammatone filterbank and the modulation filterbank. 
This is the speech-to-reverberation modulation ratio (SRMR) metric in which speech energy at low modulation frequencies related to speech information is separated from noise and distortion components on high modulation frequencies. 

To incorporate characteristics of a human auditory filter, \cite{Yamamoto2019} extended sEPSM using a dynamic compressive gammachirp filterbank (dcGC-FB) \citep{Irino2006}, in which the level-dependent frequency selectivity and gain of the auditory filter were reasonably determined by the data obtained from psychoacoustic masking experiments \citep{Patterson2003}. 
For OIMs, it is important to introduce the appropriate level dependency to incorporate the well-known fundamental knowledge that speech intelligibility is lower as sound level decreases and that peripheral hearing loss decreases the intelligibility. Most of the existing OIMs, except for HASPI, use linear frequency analysis, which does not account for this factor. 
This model is referred to as dcGC-sEPSM, which predicted the human results of the Wiener filtering more accurately than the original sEPSM \citep{Jorgensen2011}, CSII \citep{Kates2005}, STOI \citep{Taal2011}, and HASPI \citep{Kates2014}. 

The $\mathrm{SNR_{env}}$-based OIMs (i.e., sESPM and dcGC-sEPSM) present fundamental limitations. As shown in Fig. \ref{fig:OIM}(a), the $\mathrm{SNR_{env}}$-based OIMs also require the residual noise ($\tilde{N}$), estimated by a speech-enhancement algorithm.
The definition of the residual noise was, however, not entirely clarified in the original article \citep{Jorgensen2011}.
This is an issue for most speech-enhancement algorithms, including  recent non-linear speech-enhancement algorithms \citep{Fujimoto2012,Weninger2014,Smaragdis2017}, because there are different techniques to estimate the residual noise as described in \ref{app:calc_residual_noise}. Therefore, the $\mathrm{SNR_{env}}$-based OIM approaches are restricted to speech enhancement algorithms (i.e., SS), which can estimate the residual noise uniquely and properly. 

In many applications, it is preferable to use non-intrusive OIMs that do not use a reference signal.  Yet, the prediction accuracy for speech-enhancement algorithms is still worse than that of intrusive OIMs \citep{Falk2015}. A current alternative and practical choice is to use a clean signal as a reference, as shown in Fig. \ref{fig:OIM}(b). 
As described above, there are several correlation-based OIMs, such as STOI, ESTOI, and HASPI. 

The objective of this study is to develop a new OIM, based on a reliable level-dependent auditory filterbank. 
The developed OIM is designed to predict the speech intelligibility for NH listeners at moderate sound pressure levels (SPLs) with higher or comparable performance levels to the current OIM. The new OIM should have advantageous aspects and serve as a reliable base for future studies on speech intelligibility prediction for HI listeners or in conditions with a wide range of SPLs. 

\subsection{Proposed methods}
\label{sec:Intro_new}

\begin{figure}[tb]
  \centerline{\includegraphics[width=0.85\columnwidth]{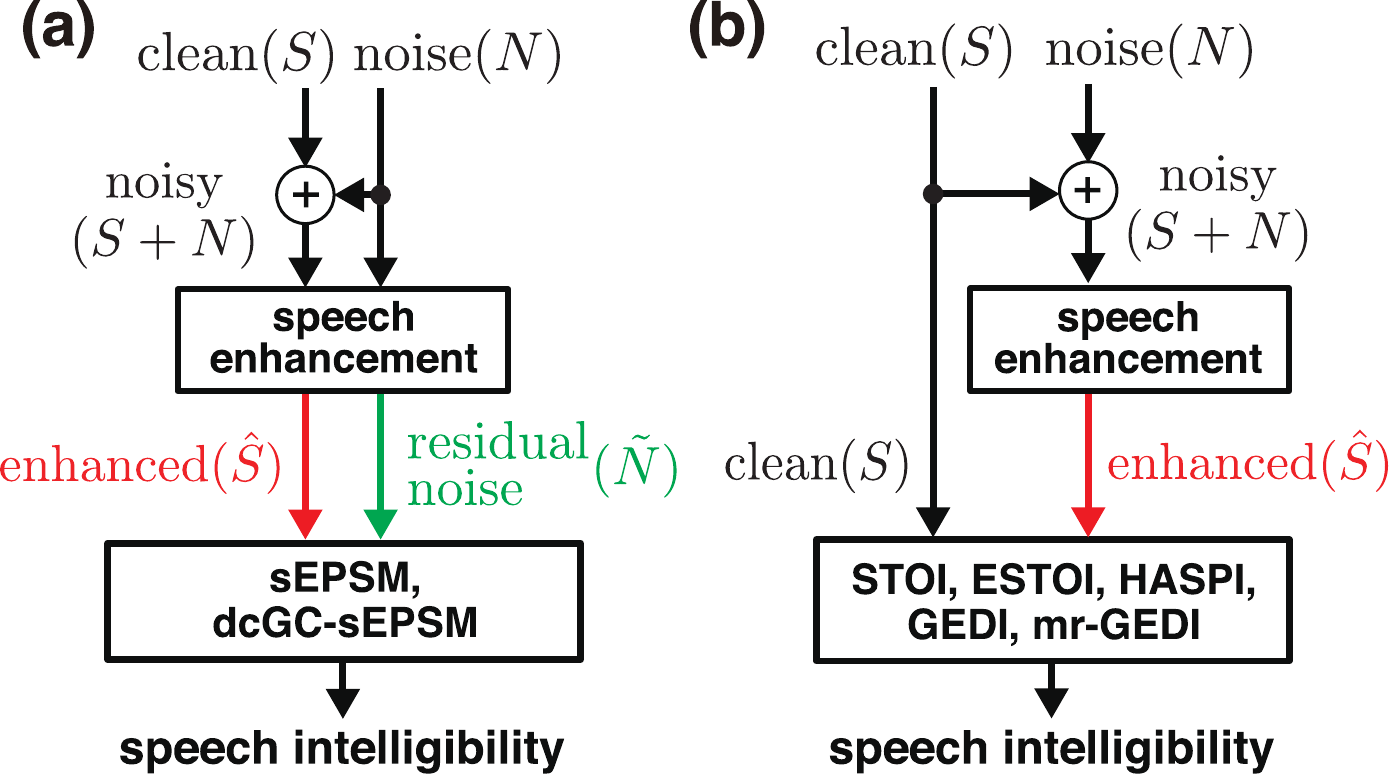}}
  \caption{Two types of OIMs requiring a reference signal for speech intelligibility prediction. (a) sEPSM and dcGC-sEPSM use residual noise ($\tilde{N}$) as a reference. (b) GEDI, mr-GEDI, and other major models use clean speech ($S$) as a reference.} 
  \label{fig:OIM}
\end{figure}

In this paper, we demonstrate a new OIM called ``gammachirp envelope distortion index (GEDI),'' which calculates the signal-to-distortion ratio in the envelope domain ($\mathrm{SDR_{env}}$) and uses clean speech ($S$) as the reference signal, as shown in Fig. \ref{fig:OIM}(b). The internal representations in the proposed model are similar to those of dcGC-sEPSM and original sEPSM, which use $\mathrm{SNR_{env}}$. GEDI was initially proposed by \cite{Yamamoto2017} and evaluated under pink background-noise conditions. After preliminary experiments, GEDI was extended to include a weighting function to compensate for the envelope power across auditory filter channels. The unreported effect is presented here.
Extended experiments were also performed to predict speech intelligibility under non-stationary, babble-noise conditions. 
The poor results urged us extend GEDI to a multi-resolutional version to improve predictability \citep{Yamamoto2018}. 

In this paper, we thoroughly explain GEDI and mr-GEDI and perform experiments using speech intelligibility predictions.
The prediction results of GEDI, STOI \citep{Taal2011}, ESTOI \citep{Jensen2016}, and HASPI \citep{Kates2015} are compared to human results by using speech materials produced by two speech-enhancement algorithms under pink- and babble-noise conditions. 

In Section 2, we provide an overview of GEDI and mr-GEDI. In Sections 3 and 4, we describe the speech materials and experimental conditions of the evaluation, respectively. In Section 5, prediction results and human results are compared. In Section 6, some aspects of model development are discussed.

\section{Proposed OIMs: GEDI and mr-GEDI}
\label{sec:Proposed}

\subsection{GEDI}
\label{sec:GEDI}

\begin{figure}[htbp]
  \centerline{\includegraphics[width=0.6\columnwidth]{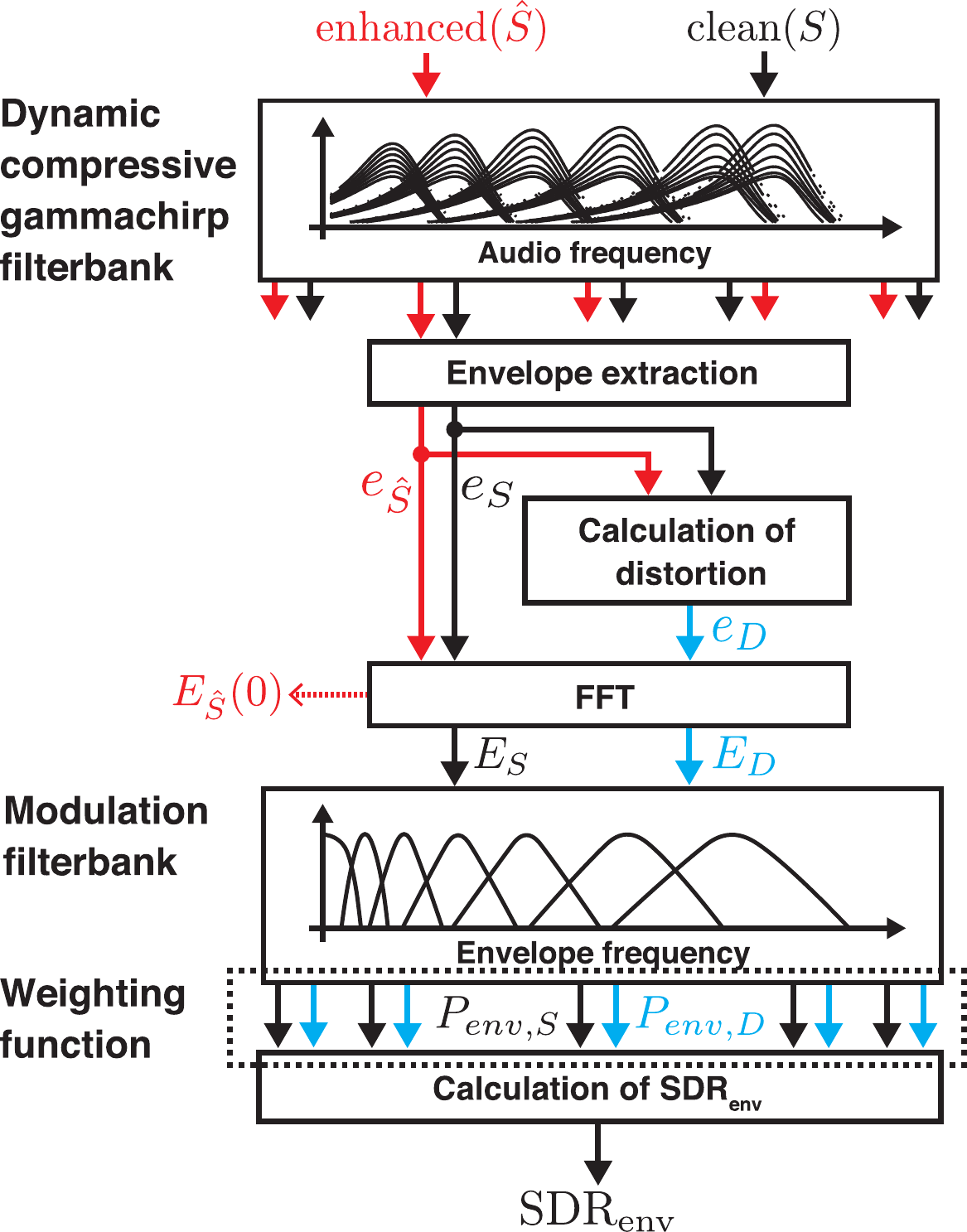}}
  \caption{Block diagram of GEDI. }
  \label{fig:Overview_GEDI}
\end{figure}

Figure\,\ref{fig:Overview_GEDI} shows a block diagram of GEDI. The input sounds to GEDI include enhanced speech ($\hat{S}$) and clean speech ($S$). The objective of GEDI is to calculate the distortion between the temporal envelopes of the clean and enhanced speech from the outputs of an auditory filterbank. We hypothesize that speech intelligibility becomes increasingly degraded as the temporal envelopes of the enhanced speech diverge from those of clean speech. 

\subsubsection{Auditory filterbank}
\label{ssec:dcGC} 
The first stage is an auditory spectral analysis using dcGC-FB \footnote{MATLAB code for the dcGC-FB is available in the GitHub repository \citep{GitHub_GCFB2019}. 
} \citep{Irino2006}, which has a number of 100 channels equally spaced along the ${\rm ERB_N}$-number \citep{Moore2013} and covers the speech range between 100 and 6,000 Hz. This is the same as in with dcGC-sEPSM. The merit of dcGC-FB is that parameter values can also be estimated from psychoacoustic experiments of elderly listeners \citep{Matsui2016} and NH listeners \citep{Patterson2003}. Although NH parameters are used in this paper, they can be extended in that direction in the future.

Figures\,\ref{fig:Envelope}(a) and (b) show examples of auditory spectrograms of enhanced ($\hat{S}$) and clean-speech signals ($S$). The auditory filter changes the gain and bandwidth in accordance with the input level. Therefore, the dcGC-FB is carefully set to correspond to the SPL used for subjective listening experiments. With GEDI, the reference signal ($S$) and the test signal ($\hat{S}$) are normalized to have the same SPL. 

\subsubsection{Distortion in the temporal envelope domain}
\label{ssec:Envelope}
\begin{figure}[tb]
  \centerline{\includegraphics[width=0.95\columnwidth]{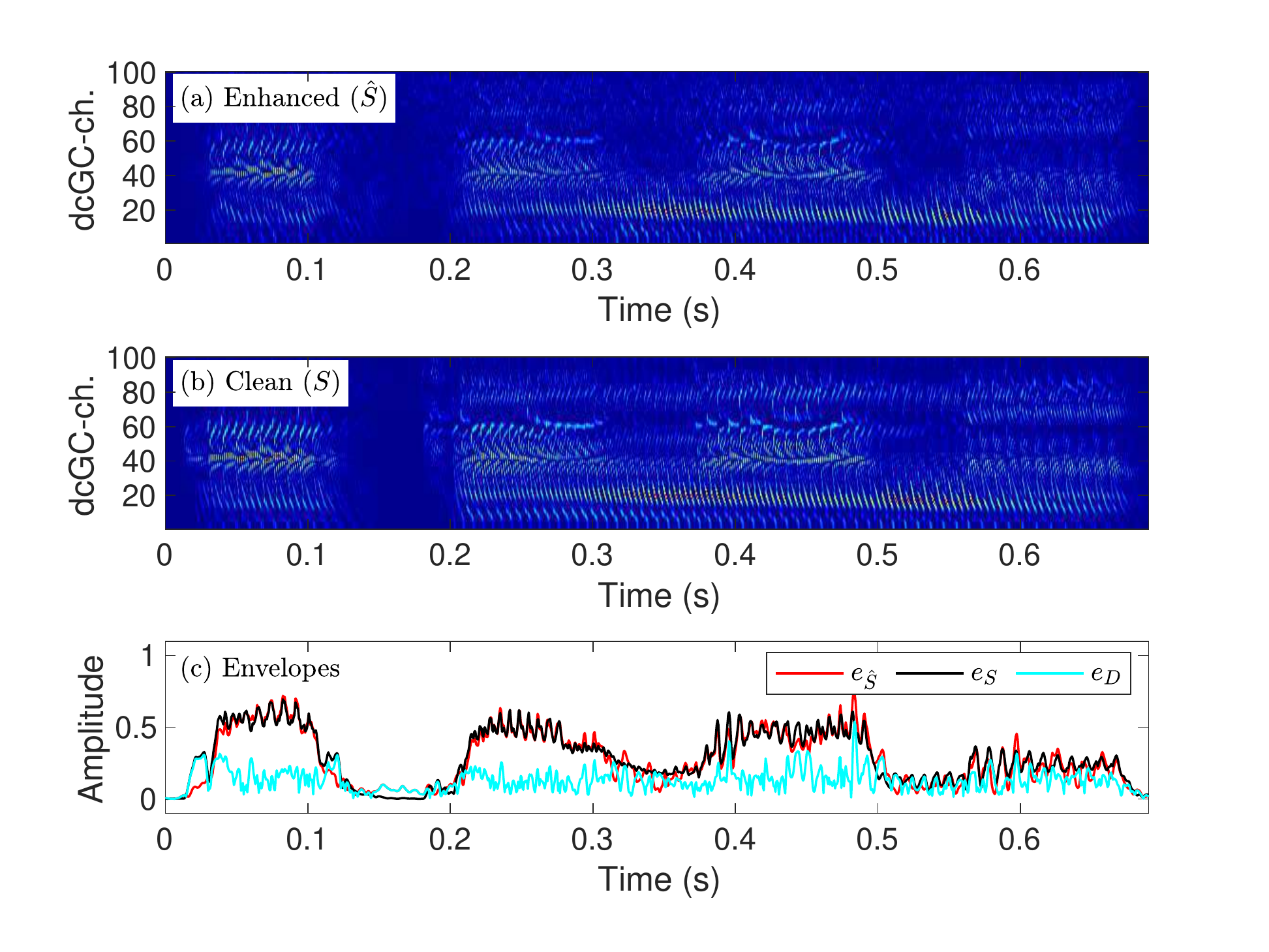}}
  \vspace{-10pt}
  \caption{Example of auditory spectrograms of a word, ``/akagane/,'' analyzed by the dcGC-FB. (a) Enhanced speech ($\hat{S}$), (b) clean speech ($S$), and (c) temporal envelopes ($e_{S}$ and $e_{\hat{S}}$) and envelope distortion ($e_{D}$) calculated from outputs from the 40-th dcGC filter.}
  \label{fig:Envelope}
\end{figure}

The temporal envelopes of the enhanced ($e_{\hat{S}}$) and clean speech ($e_{S}$) are calculated from the output of the individual auditory filter using a Hilbert transform and a low-pass filter with a cutoff frequency of 150\,Hz. The absolute difference between the two power envelopes is calculated to determine the temporal ``envelope distortion ($e_D$),'' as follows:
\begin{equation}
	e_{D,i} (n) = \bigl(|\{e_{S,i}(n)\}^{p}-\{e_{\hat{S},i}(n)\}^{p}|\bigr)^{1/p},
	\label{eq:EnvDist}
\end{equation}
where $i \{i | 1\le i \le I\}$ is the index of the dcGC-FB channel, $I = 100$ is the total number, and $n$ is the sample number of the temporal envelope. 
Here, $p$ is a constant; we set this number as $p = 2$. Thus, the envelope distortion, $e_D$, represents these differences as absolute values. Figure\,\ref{fig:Envelope}(c) shows an example of envelopes, $e_{S}$ and $e_{\hat{S}}$, and distortion, $e_D$, calculated using Eq. \ref{eq:EnvDist}. The use of the enhancement algorithms causes the envelope of the enhanced speech to either be emphasized or degraded relative to that of clean speech. The temporal envelope of the enhanced speech differs from that of clean speech. 
A new working hypothesis introduced by 
GEDI is that the envelope distortion, calculated by Eq.\ \ref{eq:EnvDist}, is negatively correlated to speech intelligibility \citep{Yamamoto2017,Yamamoto2018}.
That is, the relative distortion power is strongly related to speech intelligibility as speech intelligibility decreases when the power of the envelope distortion increases and vice versa. 

\subsubsection{$\mathit{SDR}$ in the envelope modulation domain}
\label{ssec:GEDI_SDRenv}
The modulation spectra of the envelope distortion ($e_D$) and the envelope of the clean speech ($e_S$) are calculated using the fast Fourier transform (FFT). 
A bank of modulation filters, defined in envelope frequency domain ($f_{env}$), is applied to the spectra. 
There are seven modulation filters whose power spectra are $W_{f_{env}^c}(f_{env})$ for the modulation center frequency of $f_{env}^c$, as illustrated in Figure\,\ref{fig:Overview_GEDI} and described in previous studies \citep{Jorgensen2011,Yamamoto2019}. 
The envelope power at the output of the modulation filter is calculated as
\begin{equation}
	P_{env,*} 
		= \frac{1}{E_{\hat{S}}(0)^2}
			\int_{f_{env} > 0}^{\infty}
				{|E_{*}(f_{env})|^2 W_{f_{env}^c}(f_{env})~df_{env}},\\
	\label{eq:GEDI_Penv}
\end{equation}
where the asterisk ($*$) represents either $S$ or $D$, and $E_{\hat{S}}(0)$ represents the 0-th order coefficient of the FFT (i.e., the direct-current (DC) component of the temporal envelope). 
\cite{Yamamoto2019} reported that normalization of the $E_{\hat{S}}(0)$ in dcGC-sEPSM was effective for speech intelligibility prediction of enhanced speech. 
The normalization has been inherited by GEDI which has the same filterbank structure as dcGC-sEPSM. 
The common denominator preserves the difference of levels between modulation components of the reference and enhanced speech sounds. In the original sEPSM \citep{Jorgensen2011}, it was assumed that there was internal noise in the modulation domain to restrict the lower limit of $P_{env,*}$. The formula, 
\begin{equation}
    P_{env,*} = \max(P_{env,*},0.01),
\end{equation}
was also used in GEDI. The total number of $P_{env,*}$ is 700, because the total number of the dcGC-FB channels, $I$, is 100, and the total number of the modulation filters, $J$, is 7.

The SDR in the modulation frequency domain (${\rm SDR_{env}}$) is calculated as the ratio of the modulation power spectra of clean speech, $P_{env, S}$, to the distortion, $P_{env,D}$. The individual, ${\rm SDR}_{{\rm env},j}$, for modulation filter channel, $j$, is defined as the ratio of the powers summed across the dcGC-FB channel, $i$, which can be written as
\begin{equation}
	{\rm SDR}_{{\rm env},j}^W = \frac{ \sum_{i=1}^{100} W_{i} \cdot P_{env,S,i,j} }{\sum_{i=1}^{100} W_{i} \cdot {P}_{env,D,i,j}}, 
	\label{eq:GEDI_SDRenv_k_Weight}
\end{equation}
where $W_{i}$ is a weighting function described in next subsection.
The total, ${\rm SDR}_{{\rm env}}$ can be calculated as
\begin{equation}
	{\rm SDR}_{{\rm env}} 
	=  \sqrt{\sum_{j=1}^J\bigl({\rm SDR}_{{\rm env},j}^W\bigr)^2}, 
	\label{eq:GEDI_Total_SDRenv_dcGC}
\end{equation}
where $j$ is the index number of the modulation filter, $\{j | 1\le j \le J\}$.

\subsubsection{$\mathit{SDR_{env}}$ with a weighting function}
\label{ssec:GEDI_weighted}

The weighting function, $W_i$, in Eq.\,\ref{eq:GEDI_SDRenv_k_Weight} was introduced to improve speech intelligibility prediction (see \ref{app:Effect_of_Wi} for detail).  
The envelope power, $|E_{*}(f_{env})|^2$, in Eq. \ref{eq:GEDI_Penv} is proportional to the output power of the dcGC-FB, because it is linearly derived by the FFT and the modulation filterbank, as shown in Fig. \ref{fig:Overview_GEDI}. 
The output power of the individual auditory filter is proportional to the rectangular bandwidth, $\mathrm{ERB_N}$, \citep{Moore2013} defined as
\begin{equation}
    \mathrm{ERB_N}(f) = 24.7\bigl(\frac{4.37f}{1,000} + 1 \bigr), 
    \label{eq:ERB_N}
\end{equation}
where $f$ is the filter-center frequency in Hz, and the bandwidth is roughly proportional to $f$ above 500\,Hz.
Therefore, the average value of $|E_{*}(f_{env})|^2$ increases with the filter frequency. Moreover, the auditory filters in the dcGC-FB are distributed densely along the frequency axis with considerable overlapping, as described in Section \ref{ssec:dcGC}.

In the series of studies \citep{Yamamoto2017,Yamamoto2018}, 
claimed that the evaluation of speech spectrum uniformly on an $\mathrm{ERB_N}$-number axis provided better results.
Thus, the frequency-dependent level difference must be compensated by a weighting function, $W_i$, which is inversely proportional to the bandwidth of the filter at frequency $f_i$, as
\begin{equation}
	W_i = \frac{\mathrm{ERB_{N}(1,000)}}{\mathrm{ERB_N} (f_i)}.
	\label{eq:WeightingFunction}
\end{equation}
Note that $W_i$ is normalized by $\mathrm{ERB_N(1,000)}$ at 1,000 Hz, which is near the center of dcGC filter frequencies.

\subsubsection{Transformation to speech intelligibility}
\label{Sec-02_GEDI_ssec:CalSpIntel}

The following procedure is the same as the one used by the sEPSM algorithm \citep{Jorgensen2011,Yamamoto2019}, except that $\rm SDR_{env}$ is used instead of $\rm SNR_{env}$. $\rm SDR_{env}$ is converted to a sensitivity index, $d^{\prime}$, of an ``ideal observer'' with
\begin{equation}
	d^{\prime} = k \cdot \sqrt{\rm SDR_{env}},
	\label{Sec-02_eq:dprime}
\end{equation}
where $k$ is a constant parameter determined empirically and described subsequently. 
In practice, they can be tuned such that the predicted speech intelligibility scores for the unprocessed noisy speech sounds approximately coincide with those of the human subjective score. Speech intelligibility as percentage correct, $I_{predict}$, is predicted from index $d^{\prime}$ using a multiple-alternative forced choice (mAFC) model \citep{Green1988} with an unequal-variance Gaussian model \citep{Mickes2007}, and can be written as
\begin{equation}
	I_{predict}^{(d^{\prime})}= 100 \cdot \Phi \left(\frac{d'- \mu_N}{\sqrt{\sigma_S^2 + \sigma_N^2}}\right),
	\label{Sec-02_eq:GEDI_idealobserver}
\end{equation}
where $\Phi$ denotes the cumulative normal distribution. The values of $\mu_N$ and $\sigma_N$ are determined by response-set size, $m$. 
The value of $m$ is fixed at 20,000 as reported in \citep{Yamamoto2019}. The values of $\mu_N $ and $\sigma_N $ are determined by Eqs.\,A1 and A2 from the Appendix of \citep{Jorgensen2011}. The derived values are:  $\mu_N = 4.0389$ and $\sigma_N = 0.3297$.  These values are related to the ability of an ideal observer and are not affected by experimental conditions. 
The variable of $\sigma _S$ is a parameter related to the redundancy of the speech material (e.g., meaningful sentences or monosyllables) and is determined based on the speech intelligibility experiment. The values of $\sigma _S$ and $k$ are explained in Section \ref{ssec:Eval_Obj}.

\subsection{Multi-resolution GEDI (mr-GEDI)}
\label{sec:mr-GEDI}

GEDI analyzes the modulation spectrum of the whole speech signal at once. Although it would be sufficient for stationary noise, it is difficult to separate the components of speech and non-stationary noise. Therefore, GEDI is extended to use multi-resolutional temporal frames dependent on the modulation frequency at the output of the modulation filterbank. High (low) modulation frequency characteristics are captured with short (long) frame. Such variable-frame processing has been demonstrated as beneficial for speech intelligibility predictions under non-stationary noise conditions \citep{Jorgensen2013,Rhebergen2009,Taal2011}. 

The main differences between GEDI and mr-GEDI lie in the temporal processing using IIR filters, as described in Section\,\ref{ssec:mrGEDI_modfilt}, and segmentation using different frame lengths, as described in Section\,\ref{ssec:mrGEDI_seg_Penv}.

\subsubsection{Front-end processing}
\label{ssec:proposed_aud}
\begin{figure}[t]
  \centerline{\includegraphics[width=0.6\columnwidth]{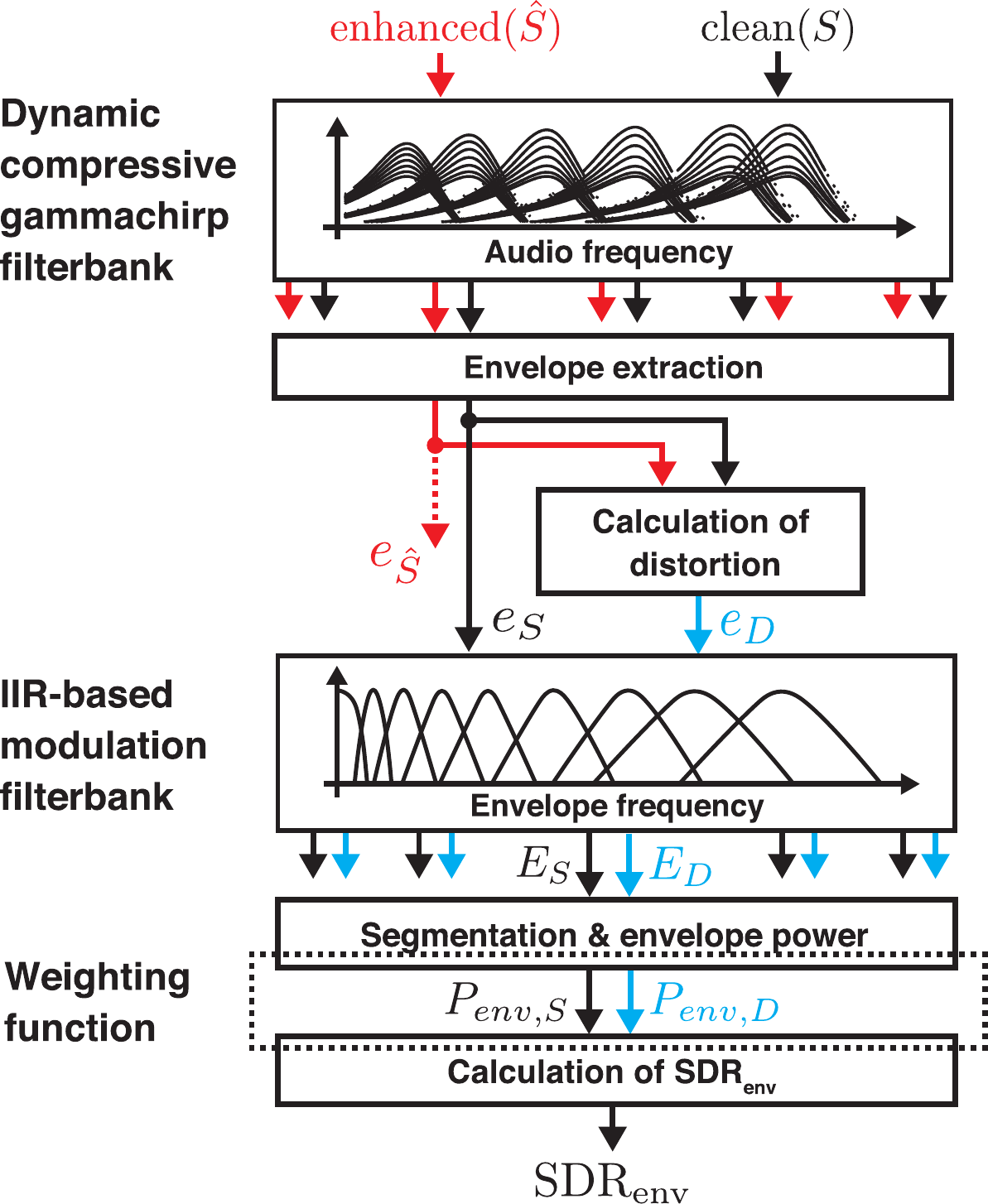}}
  \caption{Block diagram of mr-GEDI.}
\label{fig:Overview_mrGEDI}
\end{figure}

Figure \ref{fig:Overview_mrGEDI} shows a block diagram of mr-GEDI. The front-end processing, which includes the dcGC-FB, envelope extraction, and calculation of distortion, is common to the original GEDI, as shown in Fig. \ref{fig:Overview_GEDI} and as described in Sections \ref{ssec:dcGC}, \ref{ssec:Envelope}, and \ref{ssec:GEDI_SDRenv}.

\subsubsection{IIR-based modulation filterbank}
\label{ssec:mrGEDI_modfilt}
Temporal envelopes $e_{S}$ and distortion $e_D$ are filtered using an IIR-based modulation filterbank that includes a third-order low-pass modulation filter and eight second-order modulation bandpass filters. The octave-frequency space, the range, and the Q-value of the modulation filterbank are the same as in the mr-sEPSM study \citep{Jorgensen2013}. 

\subsubsection{Segmentation and envelope power}
\label{ssec:mrGEDI_seg_Penv}
The output of the $j$-th modulation filter channel, $\{j | 1\le j \le 9\}$, is segmented into multi-resolution frames using a rectangular window without overlap and is denoted as $E_{i,j}(n)$. The duration of the window equals to the inverse of the center frequency of the corresponding modulation filter  \citep{Jorgensen2013}. For example, center frequencies at 2, 4, and 8\,Hz, correspond to frame durations of 500, 250, and 125\,ms, respectively. It is 1,000\,ms for the low pass filter with the cutoff frequency of 1\,Hz. Note that the maximum frame duration becomes the signal duration when it is less than 1,000\,ms. This frame processing enables us to analyze the components with optimal resolution. The power of each frame, $P_{env}$, is calculated from the squared sum of each temporal output of the modulation filterbank:   
\begin{equation}
	P_{env, *, i, j, t}  = \frac{1}{[\overline{e_{\hat{S},i}}]^2/2} \overline{[E_{*,i,j,t}(n) - \overline{E_{*,i,j,t}}]^2},
\label{eq:mrGEDI_Penv}
\end{equation}
where the asterisk ($*$) represents components from either the clean speech, ``$S$,'' or the distortion, ``$D$.'' $t\{t | 1\le t \le T(j)\}$ is the frame index in the $j$-th modulation filter, and the bar is the average operator over the duration of the input signal. $n$ is the sample number of temporal envelopes. The denominator, $\overline{e_{\hat{S_i}}}$, in Eq.\ref{eq:mrGEDI_Penv} represents the normalization factor obtained using the DC component of the temporal envelope of the enhanced speech, $\hat{S_i}$. $P_{env,*,i,j,t}$ must be greater than $-30$\,dB (0.001 in linear terms), as suggested by \cite{Jorgensen2013}.

\subsubsection{Calculation of $\mathit{SDR_{env}}$ and speech intelligibility }
\label{ssec:mrGEDI_SDRenv}
The SDR in the temporal envelope domain (${\rm SDR_{env}}$) is calculated as the power ratio between the clean speech ($P_{env,S,i,j,t}$) and the distortion signal ($P_{env,D,i,j,t}$). The individual, ${\rm SDR}_{{\rm env},j,t}$, for modulation filter channel $j$ and frame index $t$ is defined as the ratio of the powers summed across the dcGC-FB channel, $i$, and can be written as
\begin{equation}
	{\rm SDR}_{{\rm env},j,t}^W = \frac{ \sum_{i=1}^{100} W_{i} \cdot P_{env,S,i,j,t} }{\sum_{i=1}^{100} W_{i} \cdot {P}_{env,D,i,j,t}}, 
	\label{eq:mrGEDI_SDRenv_jk}
\end{equation}
where $W_{i}$ is a weight function described in section \ref{ssec:GEDI_weighted}. The values of ${\rm SDR}_{{\rm env},j,t}^W$ are averaged over the frames $T(j)$: 
\begin{equation}
{\rm SDR}_{{\rm env,}j} =  \frac{1}{T(j)} \sum_{t=1}^{T(j)} {\rm SDR}_{{\rm env,}j,t}^W.
\label{eq:mrGEDI_Total_SDRenv_dcGC}
\end{equation}
The total ${\rm SDR}_{{\rm env}}$ is calculated by using Eq.\,\ref{eq:GEDI_Total_SDRenv_dcGC}, with $J=9$ as the number of modulation filter channels. The $\mathrm{SDR_{env}}$ in Eq.\,\ref{eq:mrGEDI_Total_SDRenv_dcGC} is transformed into speech intelligibility using Eqs.\,\ref{Sec-02_eq:dprime} and \ref{Sec-02_eq:GEDI_idealobserver}, which are the same as those used in GEDI.

\section{Speech materials for evaluation}
\label{sec:GEDI_Materials}

\subsection{Speech data}
\label{ssec:GEDI_Evaluation_Speech}
Speech sounds made by Japanese 4-mora words spoken by a male speaker (label ID: mis) from a database of familiarity-controlled word lists from 2007 (FW07) \citep{Kondo2007}, used for subjective listening experiments and objective evaluations. The database comprises several word-familiarity ranks corresponding to the degree of lexical information. Speech sounds were obtained from the set having the lowest familiarity to prevent listeners from complementing answers with guesses. The dataset contains 400 words per single familiarity, and the average duration of a 4-mora word is approximately 700\,ms. The sampling frequency of original speech sounds in the database is 48,000\,Hz. The sounds are down-sampled to 16,000\,Hz to remain consistent with the sampling frequency of speech-enhancement algorithms, used in the evaluation as described in Section \ref{ssec:GEDI_Evaluation_SE}.

\subsection{Noise conditions}
\label{ssec:GEDI_Evaluation_Noise}
Pink noise and babble noise are used for the subjective listening experiments and objective predictions. Each noise is added to the clean speech to obtain noisy speech sounds, referred to as ``unprocessed.'' The babble noise has a temporal fluctuation in power preventing the perception of individual speech. A speech-babble noise is generated from the corpus of spontaneous Japanese (CSJ) data \citep{Furui2000,Maekawa2003}, generated as follows: 8-min sections randomly extracted from each file, all superimposed as babble noise. We mixed speech signals of 32 speakers after concatenating the sentences into a single-track sound. 32 was chosen so that verbal information of individual speech would not be discerned, and that the mixed sound would not be a steady noise. 

Pink and babble noises were extracted from a random starting point before adding them to speech sounds. When making noise speech, we randomly cut out the start point from the noise, and the length is adjusted to the original speech sound. The SNR conditions range from $-6$ to $+3$\ dB in $3$-dB steps for pink-noise conditions, and from $-6$ to $+6$\ dB for babble-noise conditions. 

\subsection{Speech-enhancement algorithms}
\label{ssec:GEDI_Evaluation_SE}
In this study, we applied two speech enhancement algorithms to the unprocessed sounds. The first one was a simple SS algorithm \citep{Berouti1979}, used to ensure consistency with the method previously used to evaluate the original sEPSM method \citep{Jorgensen2011}. The second one was a Wiener filter (WF)-based algorithm. It is commonly used in various systems because of its effectiveness with low computational costs. The WF achieves speech enhancement by changing signal gains adaptively in the time-frequency bins. As the characteristics of signals enhanced by the SS and the WF are substantially different, it is essential to investigate the difference between speech intelligibility predictions when using the two methods. In our experiments, we used a pre-trained speech model (PSM) based approach \citep{Fujimoto2012} for estimating WF. The WF algorithm using the PSM is referred to as $\mathrm{WF_{PSM}}$. 

\subsubsection{Spectral subtraction}
\label{sssec:GEDI_Evaluation_SE_SS}
The amplitude spectrum of clean speech, $\hat{S}(f)$, was estimated using SS \citep{Berouti1979}, defined as:
\begin{equation}
 | \hat{S}(f) |^{2} = \left\{ \begin{array}{l}
		P_{S+N}(f)-\alpha \hat{P}_N(f)\\
		\hspace{15mm} \mathrm{when} \hspace{5mm} P_{S+N}(f) > (\alpha+\beta)\hat{P}_N(f)\\
		\beta \hat{P}_N(f)	\hspace{3.5mm}	\mathrm{otherwise}\\
	\end{array}, \right. 
    \label{eq:SpecSub}
\end{equation}
where $\hat{P}_N(f)$ represents the noise power spectrum ($N$) estimated from a non-speech segment, and $P_{S+N}(f)$ is the power spectrum of noisy speech ($S+N$). The parameter, $\alpha$, denotes the over-subtraction factor, ($\alpha \le 0$), and $\beta$ denotes the spectral flooring parameter, $(0 < \beta \ll 1)$. The over-subtraction factor, $\alpha$, for the SS was fixed at 1.0 as a reference condition for comparison with the results presented in \citep{Jorgensen2011}. We calculated the power and phase spectra using a short-time Fourier transform with a $1,024$-point Hanning window and a $50\,$\% frame shift at a sampling frequency of 16,000\,Hz. This method is referred to as ``$\rm SS^{(1.0)}$.''

\subsubsection{Wiener filter with pre-trained speech model}
\label{sssec:GEDI_Evaluation_SE_WFpsm}
The $\mathrm{WF_{PSM}}$ used in this study was estimated using a PSM of the clean speech and noise \citep{Fujimoto2009,Fujimoto2012}. The PSM is defined as a Gaussian mixture model defined in the Mel-spectrum domain using a VTS-based model combination algorithm. This algorithm can estimate the speech component of noisy speech based on the PSM, which represents the statistical distribution of the spectral features of clean speech. The PSM was trained using a large speech database comprising more than 30,000 sentences spoken by 180 speakers taken from the CSJ database \citep{Furui2000,Maekawa2003}. In this evaluation, we used a PSM with a 24-channel Mel-filterbank and set the number of Gaussian mixture components for speech and noise to 64 and 1, respectively. The WF gain applied to the noisy speech in the linear frequency domain was calculated using frequency warping from the Mel-frequency domain. 

In $\mathrm{WF_{PSM}}$, the amount of residual noise can be controlled by the Wiener gain parameter, $\mathrm{\varepsilon}$ $\{\mathrm{\varepsilon} | 0 \le \mathrm{\varepsilon} \le 1\}$. 
The residual noise increases with the value of $\mathrm{\varepsilon}$. 
The $\mathrm{WF_{PSM}}$ with $\mathrm{\varepsilon}$ values of 0, 0.1, and 0.2 are referred to as ``$\rm WF_{PSM}^{(0.0)}$,'' ``$\rm WF_{PSM}^{(0.1)}$,'' and ``$\rm WF_{PSM}^{(0.2)}$,'' respectively. We used ``$\rm WF_{PSM}^{(0.0)}$,'' ``$\rm WF_{PSM}^{(0.1)}$,'' and ``$\rm WF_{PSM}^{(0.2)}$'' models for pink-noise conditions and ``$\rm WF_{PSM}^{(0.0)}$'' and ``$\rm WF_{PSM}^{(0.2)}$'' for tests under babble noise conditions because of restrictions on the experimental condition.

\section{Evaluation conditions}
\label{sec:EvalCond}
We performed subjective human experiments to estimate the intelligibility of enhanced speech described in Section \ref{sec:GEDI_Materials}. The proposed and conventional OIMs were evaluated based on how well they predicted the human results. 
Note that the speech materials used in the experiments were different for individual subjects. Therefore, the predictions were performed for the individual materials. 

\subsection{Subjective intelligibility}
\label{sec:SbjMsr}

\subsubsection{Sound presentation}
\label{ssec:SbjMsr_Prs}

For pink noise conditions, the sounds were presented diotically via a digital-to-analog (DA) converter (Fostex, HP-A8) over headphones (Sennheiser, HD-580) at a quantization level of 24 bits and a sampling frequency of 48,000\,Hz after up-sampling from 16,000\,Hz. The level of stimulus sounds was 65\,dB in $ L_{\rm Aeq}$. Listeners were seated in a sound-attenuated room with a background noise level of approximately 26\,dB in $L_{\rm Aeq}$. For babble-noise conditions, the sounds were presented diotically via a DA converter (OPPO, HA-1) over headphones (OPPO, PM-1) at a sampling frequency of 48,000\,Hz. The stimulus sound levels were 63\,dB in ${\rm L_{Aeq}}$.

\subsubsection{Listeners}
\label{ssec:SbjMsr_Sbj}

Nine young NH listeners (four males and five females) participated in the experiments under pink-noise conditions, and fourteen (eight male and six female) participated in experiments with babble-noise conditions. Their native language was Japanese. The participants had a hearing level of less than 20\,dB between 125 and 8,000\,Hz. They participated in the experiments only after providing informed consent. The participants were instructed to write down the words they heard using ``hiragana,'' which roughly corresponds to the Japanese morae or consonant-vowel syllables. The total number of presented stimuli was 400 words, comprising a combination of speech-enhancement algorithm conditions and SNR conditions with 20 words per condition. The details are listed in Table.\,\ref{tab:Conditions}. Note that the words for each condition corresponded to a set of 20 words in FW07. The total duration of the listening test was about 1 hour. To keep the listeners' attention within a reasonable range, we restricted the maximum number of words as 400 to cover all SNR conditions and enhancement algorithms. Each subject listened to a different word set, assigned randomly to avoid bias caused by word difficulty.

\begin{table}[htbp]
\caption{The numbers of SNR conditions, processing conditions, words per condition, and words used in total for each noise condition.}
\label{tab:Conditions}
    \begin{center}
    \begin{tabular}{ccccc}
    \toprule
     & \begin{tabular}[c]{@{}c@{}} SNR\\conditions \end{tabular} & \begin{tabular}[c]{@{}c@{}} processing\\conditions \end{tabular} & \begin{tabular}[c]{@{}c@{}} words per \\a condition \end{tabular} & \begin{tabular}[c]{@{}c@{}} words \\ in total \end{tabular}\\
    \midrule
    Pink    & $4$ & $5$ & $20$ & $400$\\
    Babble  & $5$ & $4$ & $20$ & $400$\\
    \bottomrule
    \end{tabular}
    \end{center}
\end{table}

\subsection{Objective intelligibility measures}
\label{ssec:Eval_Obj}

Model evaluations were performed for the prediction of human results under the conditions arising from the use of speech-enhancement algorithms and the existence of pink and babble noise conditions. STOI measure \citep{Taal2011} was selected as a \textit{de facto} standard OIM for the evaluation of state-of-the-art speech-enhancement algorithms. ESTOI measure \citep{Jensen2016} is an extended version of STOI. Additionally, HASPI \citep{Kates2015} was selected as a competing model, because it performed better than other models in a previous study \citep{Yamamoto2019}. Note that these models, including sEPSM \citep{Jorgensen2011} and CSII \citep{Kates2005}, have been used to evaluate dcGC-sEPSM in a previous study \citep{Yamamoto2019}. Thus, sEPSM failed to predict speech intelligibility for the $\mathrm{WF_{PSM}s}$ condition, and CSII could not predict speech intelligibility for $\mathrm{SS^{(1.0)}}$. Thus, we only use STOI, ESTOI, and HASPI in this study. 

We calculated speech intelligibility scores from the same speech sounds (3,600 words: 400 words $\times$ 9 listeners) under pink-noise conditions and 5,600 words (400 words $\times$ 14 listeners) under babble-noise conditions. This is because the individual subjects listened to different sets of speech sounds. Therefore, the OIM predictions were derived for the word sets provided to the individual listeners. In the following OIMs, several parameters were tuned, depending on the speech material used in the evaluation. In this study, for a fair comparison, the parameter values were determined by a least square error (LSE) method, such that the model predictions matched the intelligibility scores of human results of speech intelligibility for unprocessed conditions of all noise conditions. The capabilities of the OIMs to predict speech intelligibility based on the individual speech-enhancement algorithms were also investigated. 

\subsubsection{Parameters of GEDI and mr-GEDI}
\label{sssec:Eval_Obj_GEDI}

In GEDI, the values of the two parameters, $k$ and $\sigma_S$, in Eqs.\,\ref{Sec-02_eq:dprime} and \ref{Sec-02_eq:GEDI_idealobserver}, must be determined. The $k$ and $\sigma_S$ were determined using the LSE method to minimize the mean-squared error of the ``unprocessed'' curves between human results and the model predictions, as described above. The optimized parameter values for GEDI and mr-GEDI are listed in the first and second rows of Table \ref{tab:ParameterValues}. 

  \begin{table} [!t]
   \begin{center}
     \caption{Parameter values for GEDI, mr-GEDI, STOI, and HASPI, as described in Sections \ref{sssec:Eval_Obj_GEDI}, \ref{ssec:Eval_Obj_STOI}, and \ref{sssec:Eval_Obj_HASPI}.
    \label{tab:ParameterValues}}
\begin{tabular}{ccc}
\toprule
 & pink noise & babble noise \\
\midrule
GEDI & $k = 1.23$, \hspace{0.5em} $\sigma_S = 1.83$ & $k = 1.26$, \hspace{0.5em} $\sigma_S = 0.60$ \\
\addlinespace[5pt]
mr-GEDI &  $k = 1.43$, \hspace{0.5em}  $\sigma_S = 1.81$ &   $k = 1.50$, \hspace{0.5em}   $\sigma_S = 0.70$ \\
\addlinespace[5pt]
STOI & $a = -6.44$, \hspace{0.5em} $b = 4.56$ & $a = -8.91$, \hspace{0.5em} $b = 5.84$ \\
\addlinespace[5pt]
 ESTOI & $a = -6.16$, \hspace{0.5em} $b = 3.39$ & $a = -9.90$, \hspace{0.5em} $b = 4.88$ \\
\addlinespace[5pt]
HASPI & \begin{tabular}[c]{@{}c@{}}$B = -10.88, \hspace{0.5em}C = 4.04,$ \\ $A_{high} = 13.32$\end{tabular} & \begin{tabular}[c]{@{}c@{}}$B = -61.36, \hspace{0.5em}C= -22.15,$ \\ $A_{high} = 93.87$\end{tabular}\\
\bottomrule
\end{tabular}
   \end{center}
  \end{table}

\subsubsection{Parameters of STOI and ESTOI}
\label{ssec:Eval_Obj_STOI}
The initial STOI process comprises one-third of the octave band analysis, envelope extraction, and calculation of the short-time correlation between the envelopes of clean and target sounds in each octave. Then, the internal index of speech intelligibility measure, $d$, is obtained by averaging the inner products between sub-band temporal envelopes \citep{Taal2011}. ESTOI \citep{Jensen2016} shares its envelope extraction with STOI. The index, $d$, is instead calculated from the average of correlation coefficients between short-time spectra across sub-bands.

The speech intelligibility index of STOI and ESTOI is derived as a percentage value using a logistic function: 
\begin{equation}
    I_{predict} = \frac{100}{1+\exp(a \cdot d + b)}, 
    \label{Sec-04-eq:SpIntel_STOI}
\end{equation}
where $a$ and $b$ are parameters (See Eq.\,8 of \citep{Taal2011} and Eq.\,10 of \citep{Jensen2016}). For a fair comparison, the parameter values were adjusted to simulate human results in an unprocessed condition in our experiments (in section \ref{sec:Results}). The used values are listed in the third and fourth rows of Table \ref{tab:ParameterValues}.

\subsubsection{Parameters of HASPI}
\label{sssec:Eval_Obj_HASPI}
HASPI was developed for speech intelligibility prediction of HI listeners using an extended version of the gammatone filterbank. This index is calculated from the normalized cross-correlation of the temporal sequences of cepstral coefficients with auditory coherence values. Speech intelligibility using HASPI is derived using a logistic function, 
\begin{equation}
    I_{predict}  = \frac{100}{1+\exp(-p)}, 
    \label{Sec-04-eq:SpIntel_HASPI}
\end{equation}
as in Eqs.\,1 and 7 in \citep{Kates2014}. The parameter, $p$, is defined as a linear combination of feature values related to the cepstral correlations ($c$) and the three levels of auditory coherence ($a_{low}$, $a_{mid}$, and $a_{high}$) with a bias component, and it can be calculated as 
\begin{equation}
	p = B + C \cdot c + 0\cdot a_{low} + 0 \cdot a_{mid} + A_{high}\cdot a_{high}.
    \vspace{-1.5pt}
\label{eq:p_HASPI}
\end{equation}
The coefficients for this feature are denoted with capital letters: $B$, $C$, and $A$. Note that coefficients $A_{low}$ and $A_{mid}$ are set to zero, as described in \cite{Kates2014}. The remaining coefficients (i.e., $B$, $C$, and $A_{high}$) were determined using the LSE method. 
The fitted parameter values are listed in Table \ref{tab:ParameterValues}.

\section{Results}
\label{sec:Results}

\subsection{Human and prediction results}
\label{ssec:Results_HumanOIMs}

\subsubsection{Pink-noise conditions}
\label{ssec:Results_Pink}

\begin{figure}[htbp]
\begin{center}
\begin{minipage}[t]{0.450\linewidth}
  \centering
  \centerline{\includegraphics[width=0.965\columnwidth]{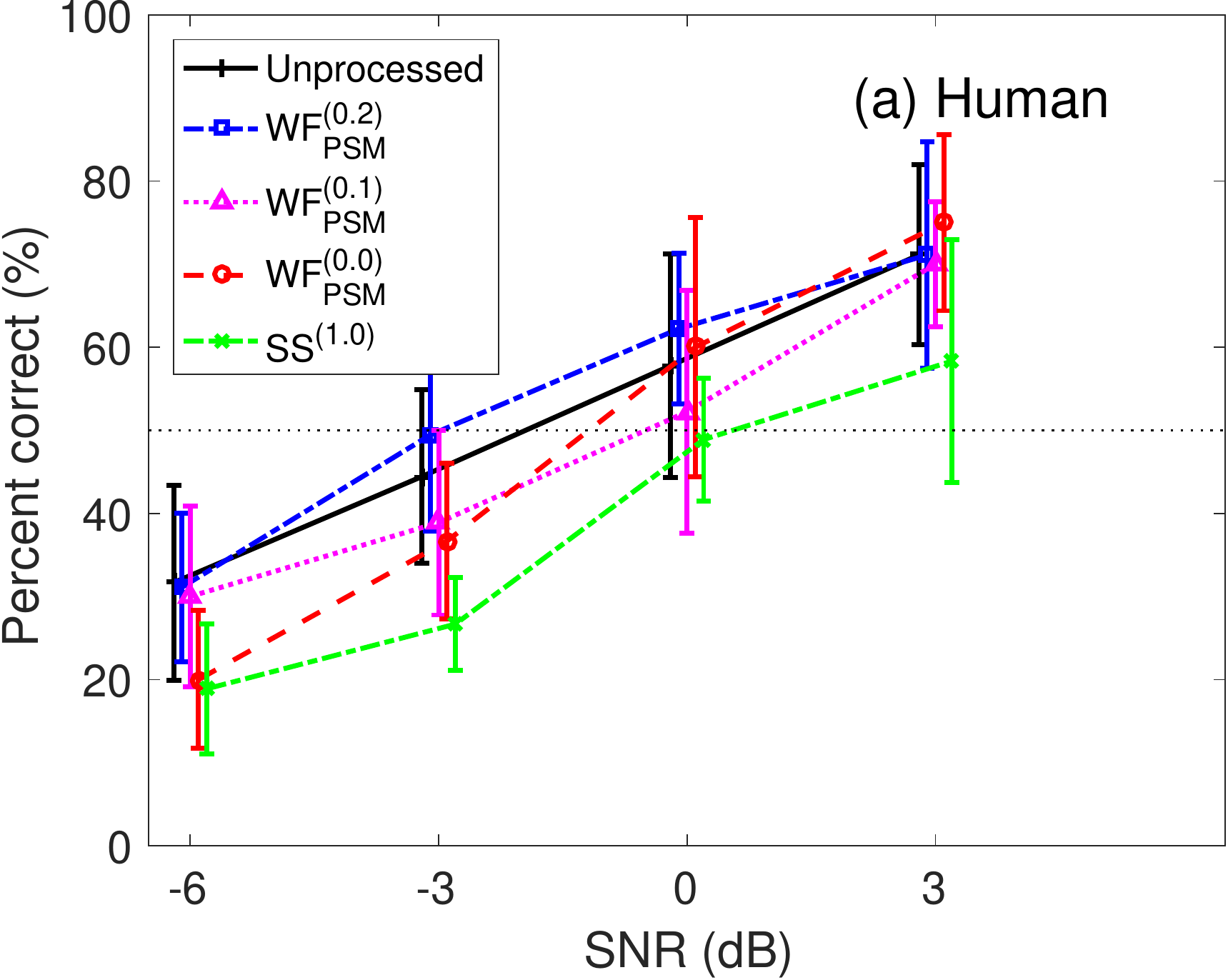}}
\end{minipage}
%
\begin{minipage}[t]{0.450\linewidth}
  \centering
  \centerline{\includegraphics[width=0.965\columnwidth]{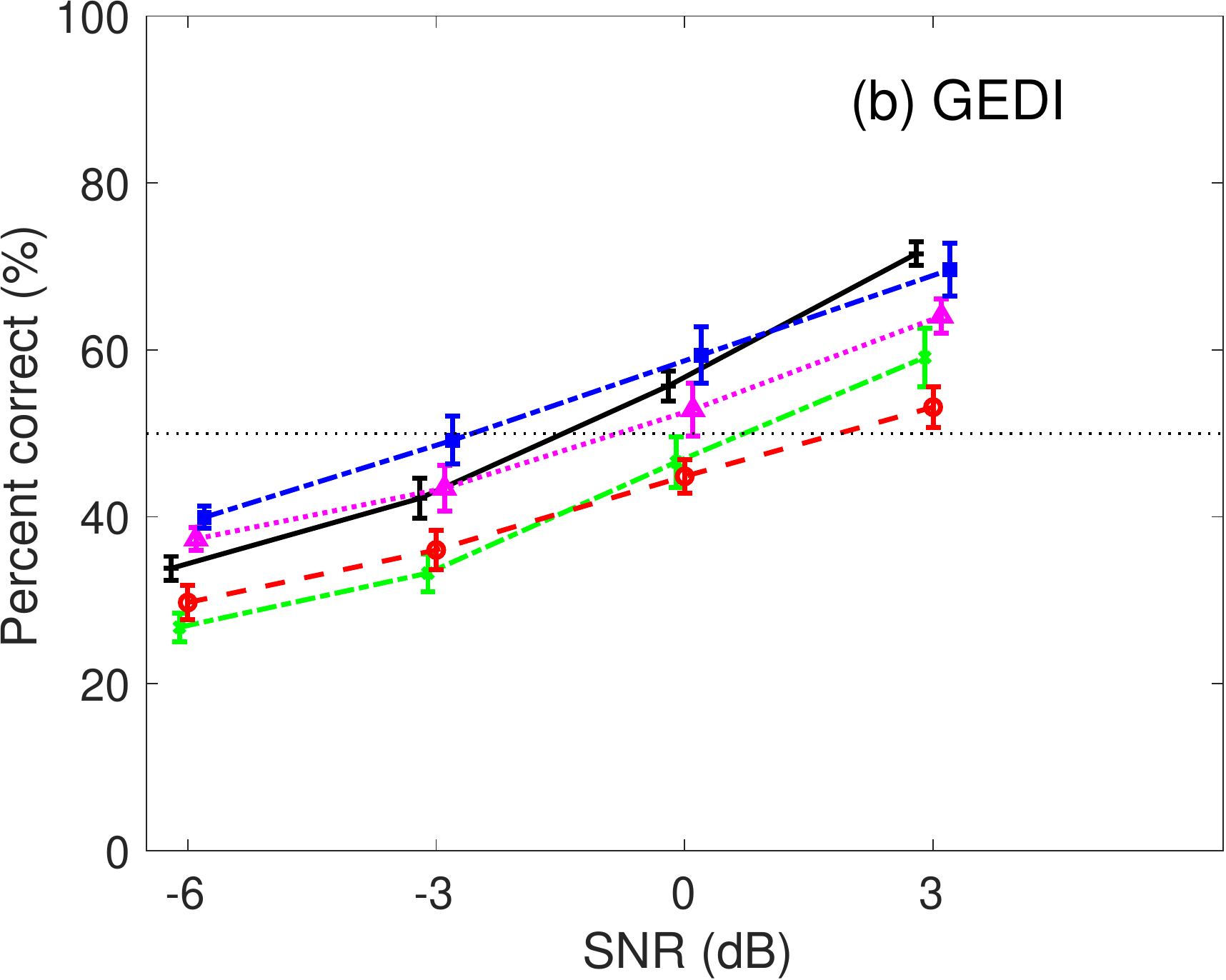}}
\end{minipage}
%
\begin{minipage}[t]{0.450\linewidth}
  \centering
  \centerline{\includegraphics[width=0.965\columnwidth]{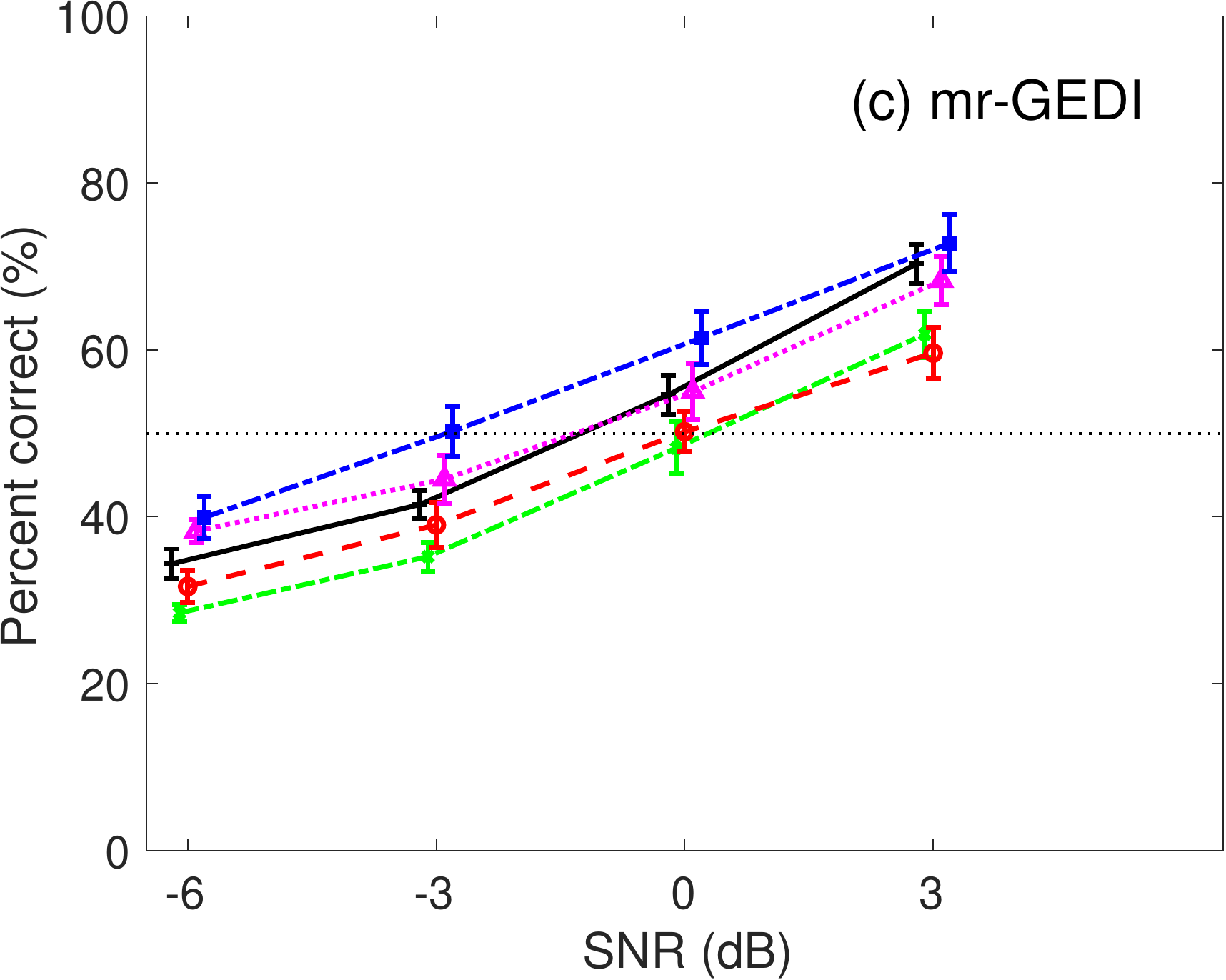}}
\end{minipage}
%
\begin{minipage}[t]{0.450\linewidth}
  \centering
  \centerline{\includegraphics[width=0.965\columnwidth]{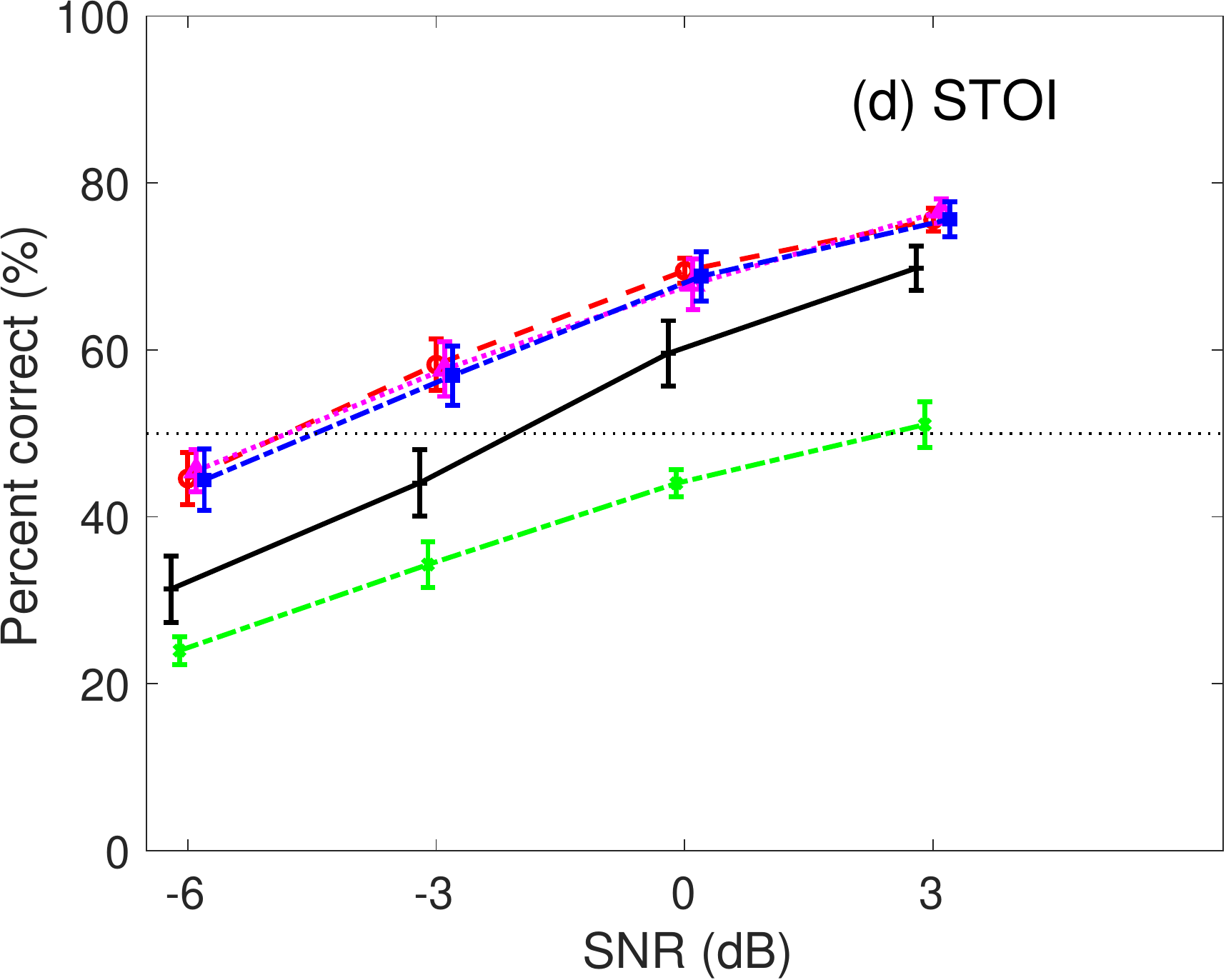}}
\end{minipage}
%
\begin{minipage}[t]{0.450\linewidth}
  \centering
  \centerline{\includegraphics[width=0.965\columnwidth]{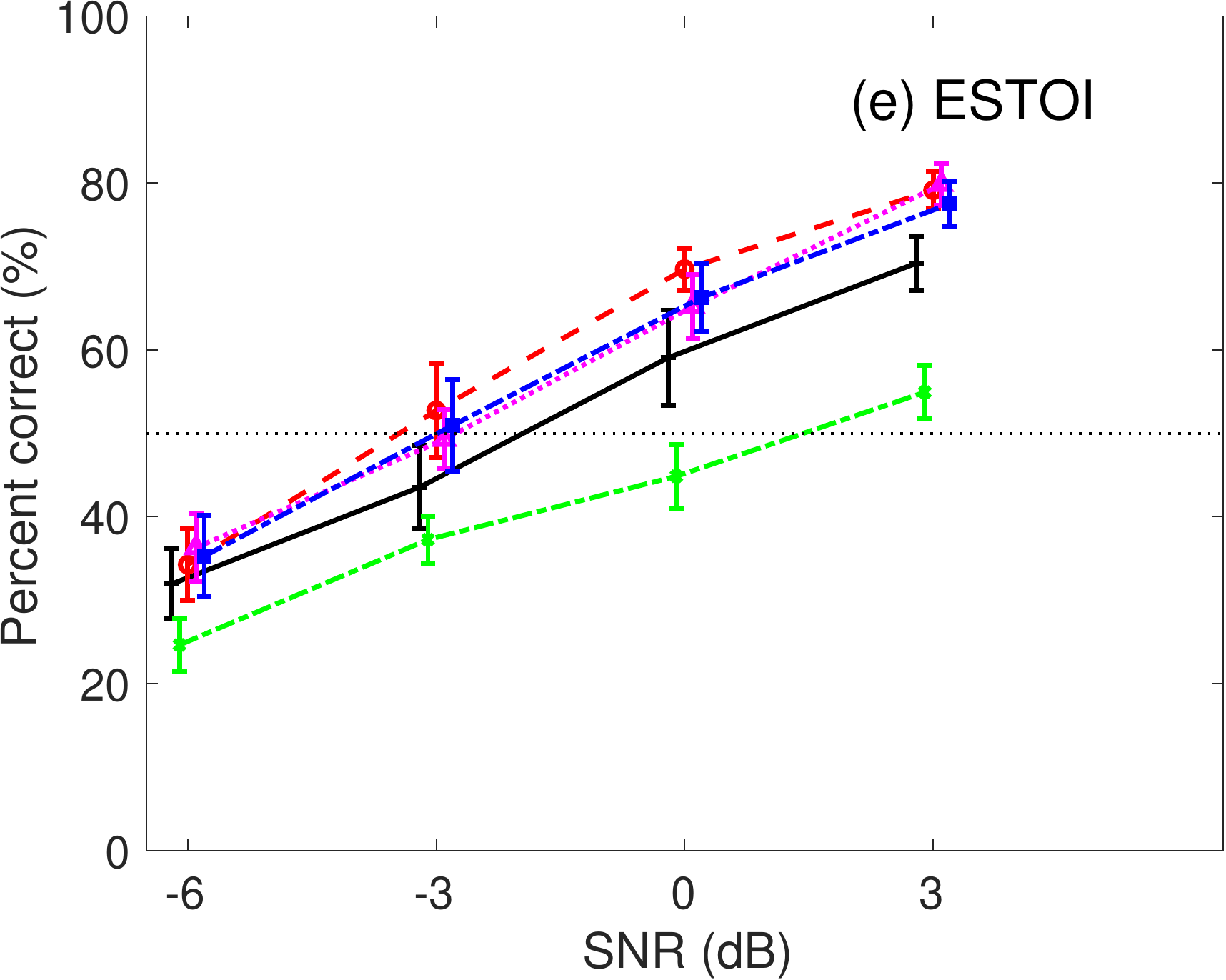}}
\end{minipage}
%
\begin{minipage}[t]{0.450\linewidth}
  \centering
  \centerline{\includegraphics[width=0.965\columnwidth]{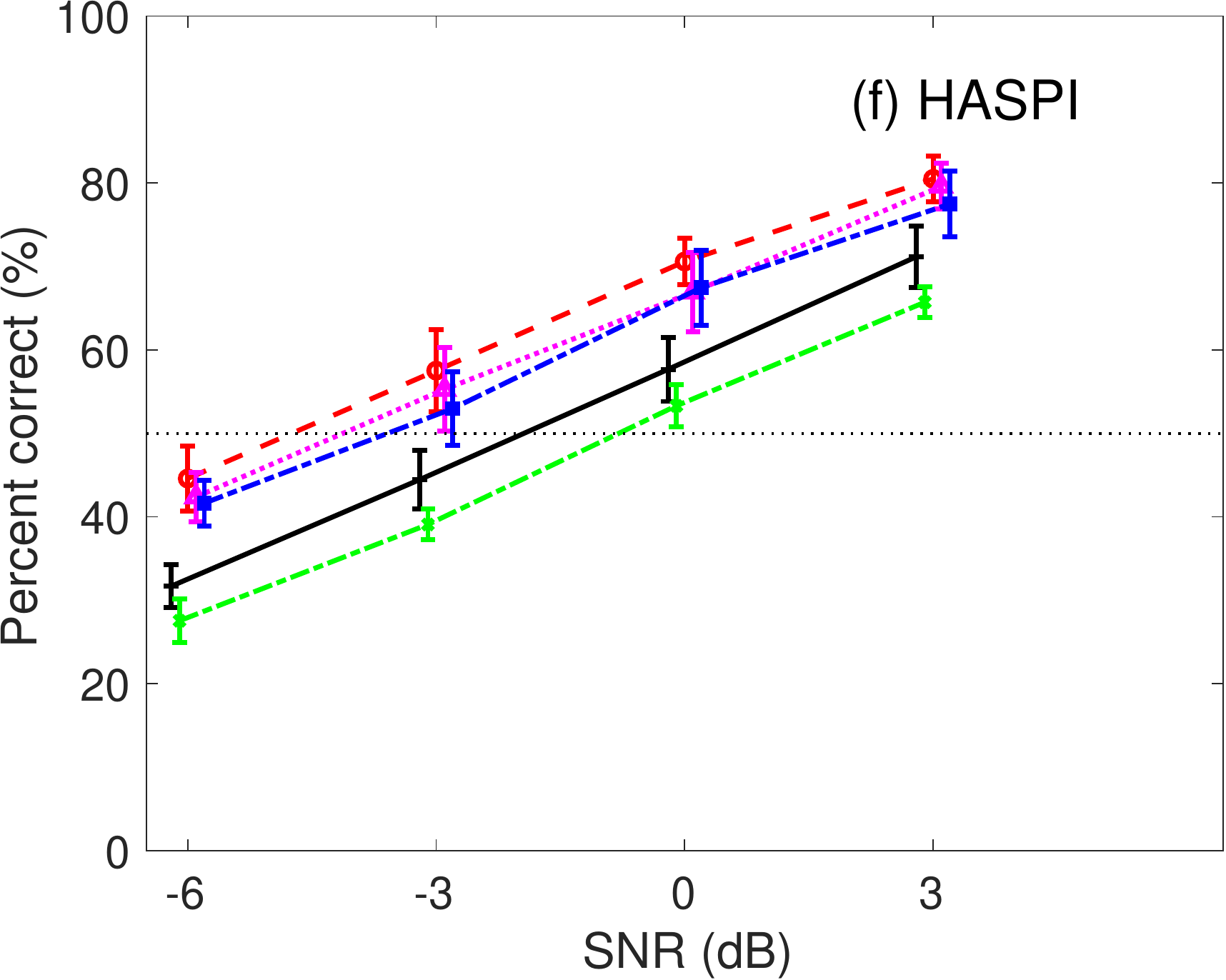}}
\end{minipage}
\caption{Results of speech intelligibility tests with pink-noise. (a) The subjective listening experiments, and the objective predictions obtained via (b) GEDI, (c) mr-GEDI, (d) STOI, (e) ESTOI, and (f) HASPI. The values and error bars represent the averages and standard deviations across the sets of speech materials.}
\label{fig:Results_PC_Pink}
\end{center}
\end{figure}

Figure\,\ref{fig:Results_PC_Pink} shows the percent correct values of speech intelligibility as a function of the speech SNR. 
Panel(a) shows the human results. 
The other panels show the model predictions of (b) GEDI, (c) mr-GEDI, (d) STOI, (e) ESTOI, and (e) HASPI. 
The speech materials for evaluation were unprocessed and enhanced sounds, which were produced by $\rm SS^{(1.0)}$ and three levels of $\mathrm{WF_{PSM}}$ (i.e., $\rm WF_{PSM}^{(0.0)}$, $\rm WF_{PSM}^{(0.1)}$, and $\rm WF_{PSM}^{(0.2)}$).  
The percentage of correct values is the averaged value across the nine noisy speech sets used for both the subjective experiments nine listeners and their objective predictions. 

For the human results (Figure\,\ref{fig:Results_PC_Pink}(a)), the speech intelligibility curves for the $\mathrm{WF_{PSM}^{(0.2)}}$ and the $\mathrm{WF_{PSM}^{(0.1)}}$ were roughly the same as the curve for the unprocessed conditions. However, the curve for the $\mathrm{SS^{(1.0)}}$ was lower than the curve for the unprocessed conditions.  
The standard deviations across listeners.  
Multiple comparison analyses with the Tukey--Kramer HSD test \citep{hsu1996multiple} ($\alpha = 0.05$) 
indicated that the speech intelligibility scores of the enhanced speech processed by $\mathrm{SS^{(1.0)}}$ were significantly lower than those of the unprocessed speech. 
There was no significant difference in any combination between the unprocessed and other enhancement methods. 

Figure\,\ref{fig:Results_PC_Pink}(b) illustrates the prediction results of GEDI. The predicted speech intelligibility curves for $\mathrm{WF_{PSM}^{(0.1)}}$ and $\mathrm{WF_{PSM}^{(0.2)}}$ are similar to the human result curves. In contrast, speech intelligibility for the $\mathrm{WF_{PSM}^{(0.0)}}$ above 0-dB SNR is lower than that of human. 

The prediction results of mr-GEDI (Figure\,\ref{fig:Results_PC_Pink}(c)) are comparable to the results of GEDI (Figure\,\ref{fig:Results_PC_Pink}(b)) as mr-GEDI has an intended design work strategy similar to GEDI with stationary noise, e.g. pink noise.

The results of STOI in Fig\,\ref{fig:Results_PC_Pink}(d) indicated  higher speech intelligibility curves for $\mathrm{WF_{PSM}^{(0.0)}}$, $\mathrm{WF_{PSM}^{(0.1)}}$, and $\mathrm{WF_{PSM}^{(0.2)}}$ compared to of the unprocessed condition. These results are inconsistent to human results. In contrast, the speech intelligibility curve for $\mathrm{SS^{(1.0)}}$ is similar to human results.

The results of ESTOI in Fig\,\ref{fig:Results_PC_Pink}(e) also show similar speech intelligibility curves as those of STOI in Fig\,\ref{fig:Results_PC_Pink}(d). Although ESTOI slightly improved the prediction of STOI, the problem of predictions for the $\mathrm{WF_{PSM}}$ family still remains.

HASPI (Figure\,\ref{fig:Results_PC_Pink}(f)) also predicted similar speech intelligibility as STOI (Figure\,\ref{fig:Results_PC_Pink}(d)) 
and ESTOI (Figure\,\ref{fig:Results_PC_Pink}(e)).
HASPI has the same problem as the speech intelligibility curves of the $\mathrm{WF_{PSM}}$ family, which are higher than the curve for the unprocessed conditions. The curve for the $\mathrm{SS^{(1.0)}}$ is within the variability of that for the human results (Figure\,\ref{fig:Results_PC_Pink}(a)).  

\subsubsection{Babble-noise conditions}
\label{ssec:Results_Babble}

\begin{figure}[htbp]
\begin{center}
\begin{minipage}[t]{0.450\linewidth}
  \centering
  \centerline{\includegraphics[width=0.965\columnwidth]{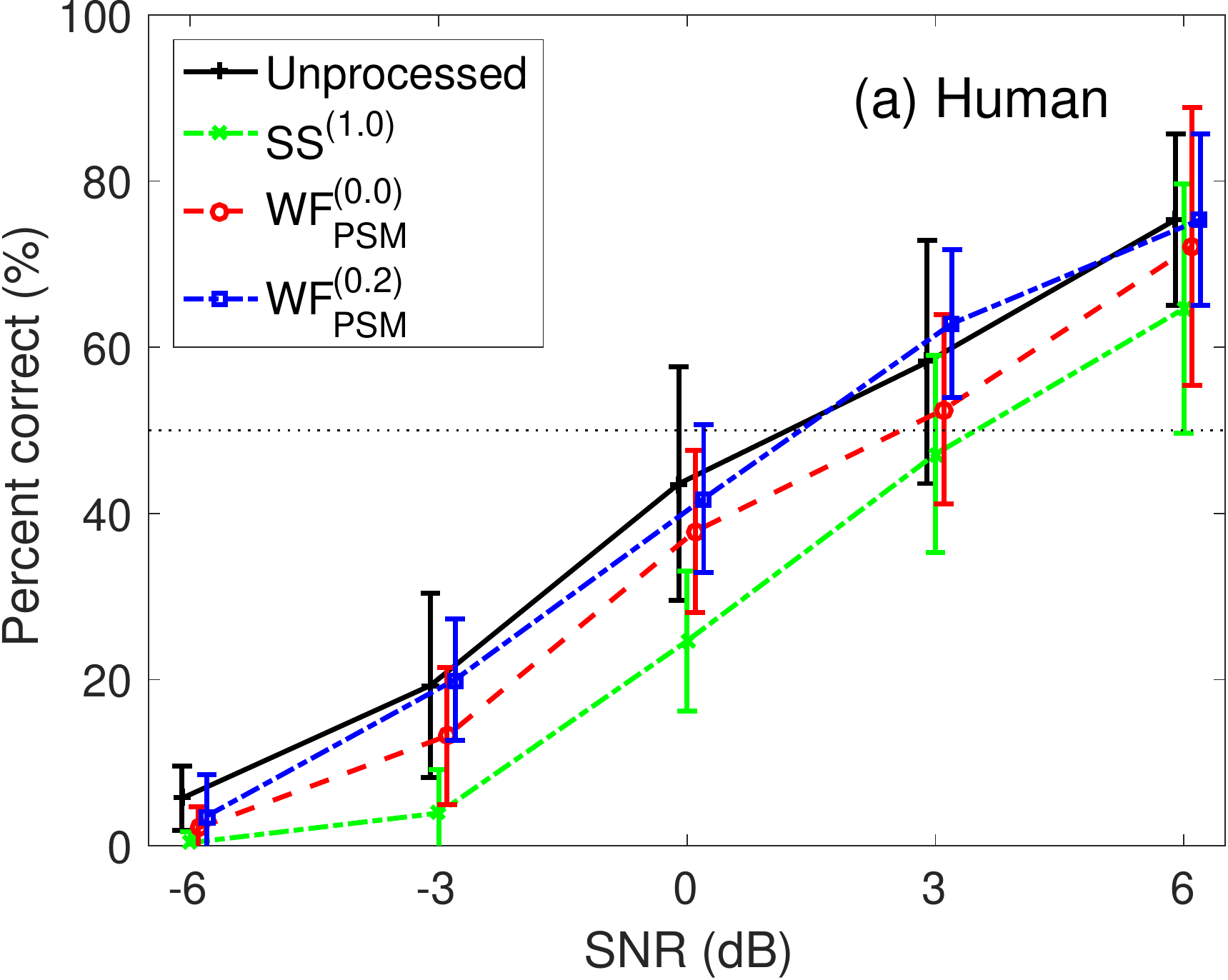}}
\end{minipage}
%
\begin{minipage}[t]{0.450\linewidth}
  \centering
  \centerline{\includegraphics[width=0.965\columnwidth]{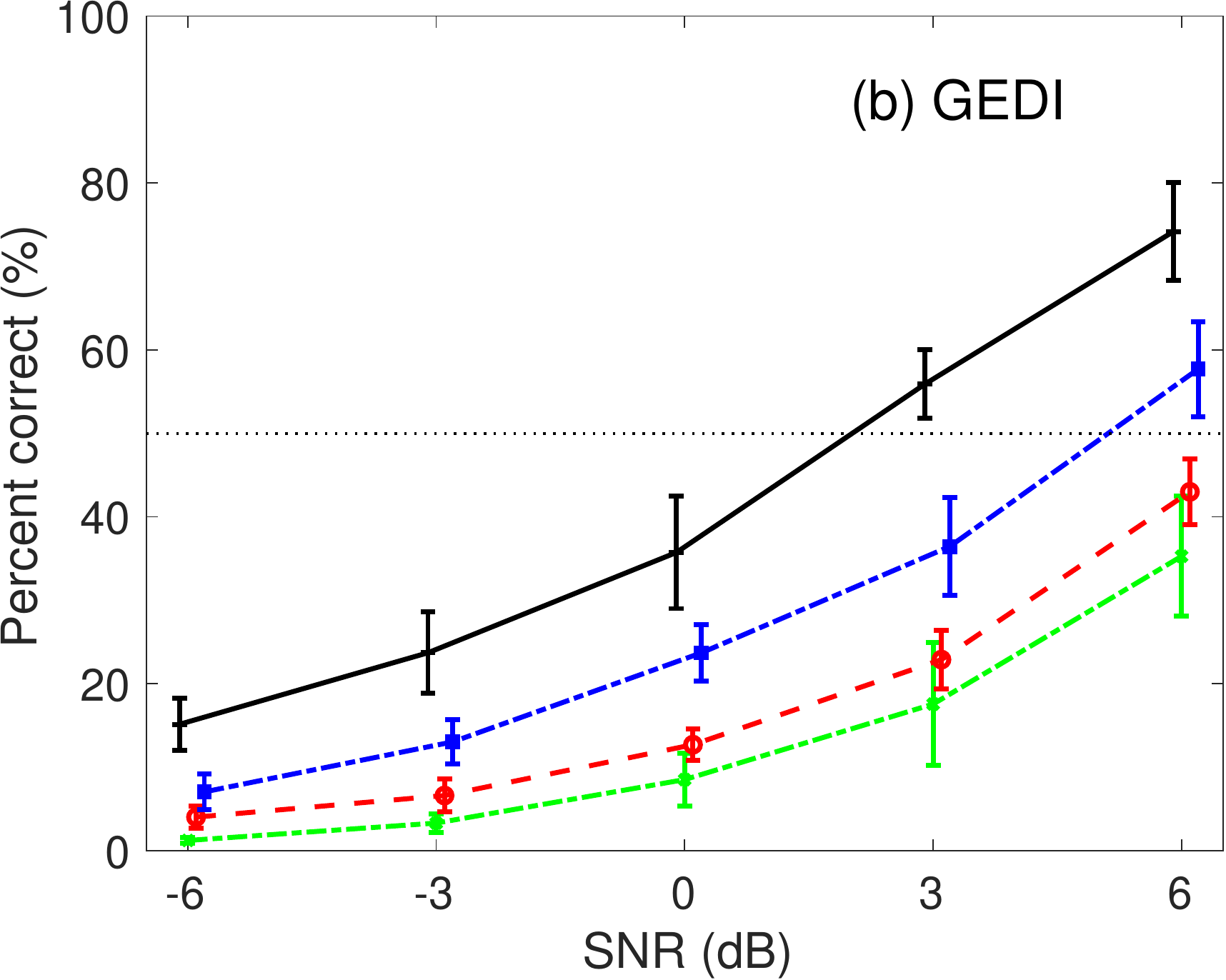}}
\end{minipage}
%
\begin{minipage}[t]{0.450\linewidth}
  \centering
  \centerline{\includegraphics[width=0.965\columnwidth]{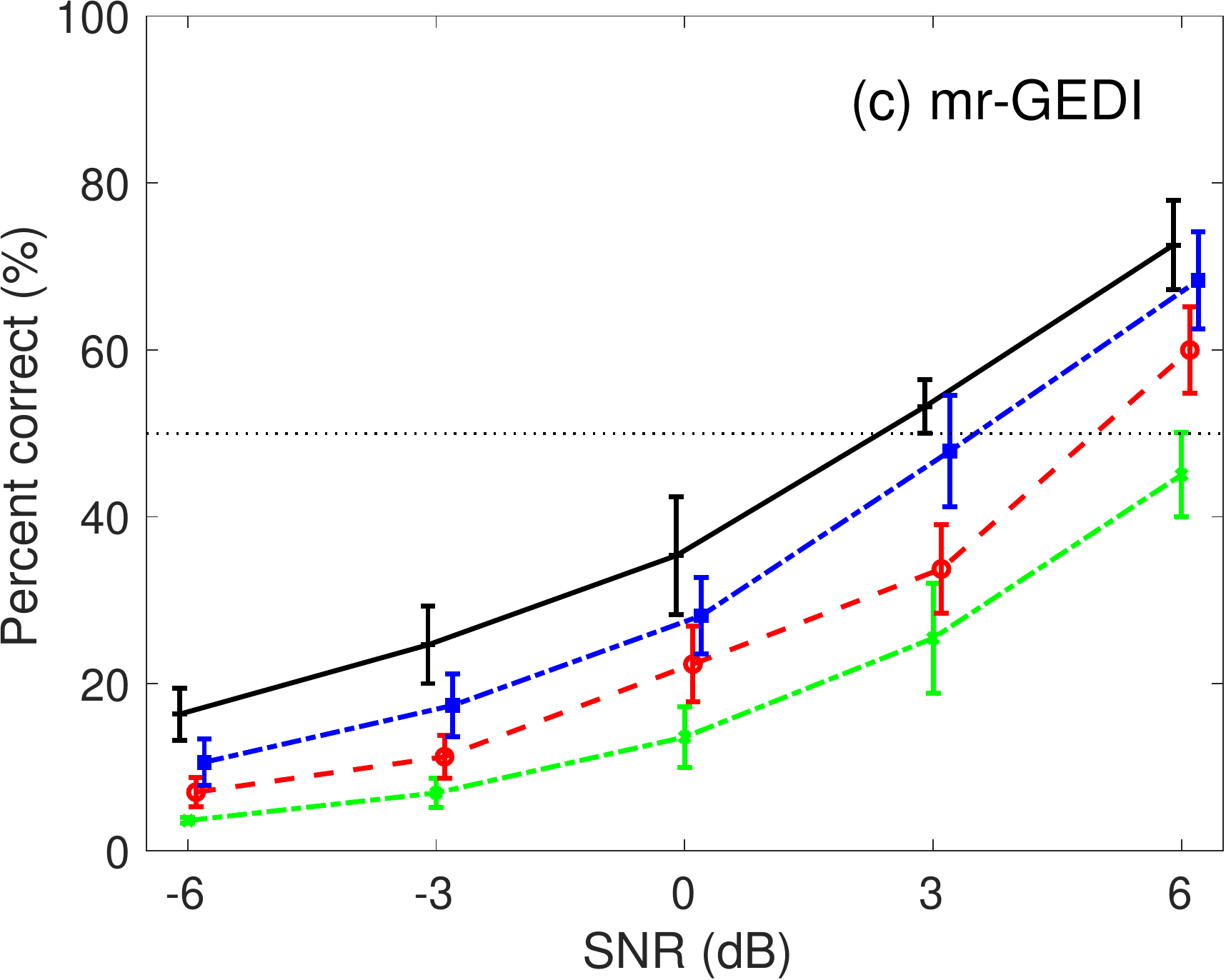}}
\end{minipage}
%
\begin{minipage}[t]{0.450\linewidth}
  \centering
  \centerline{\includegraphics[width=0.965\columnwidth]{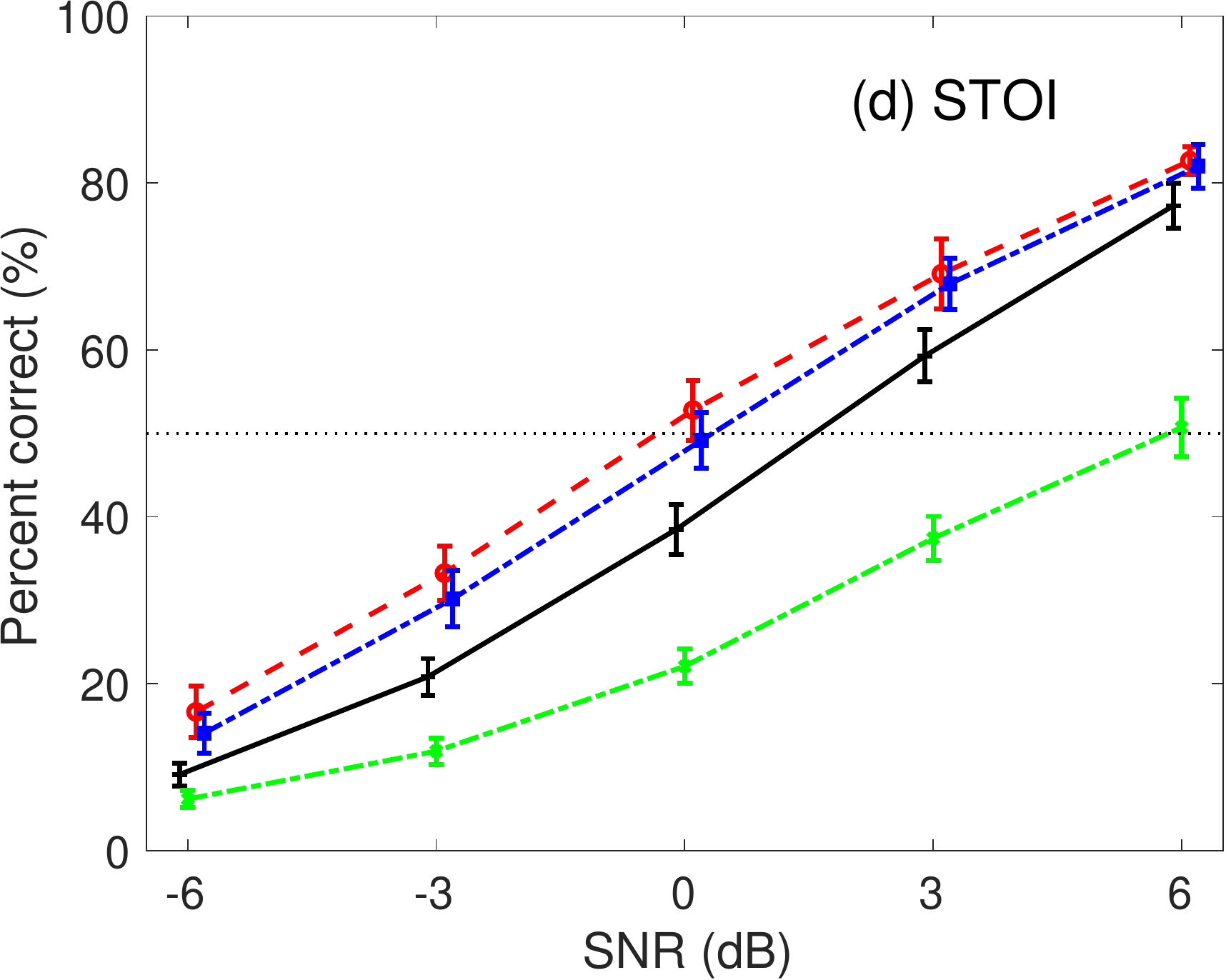}}
\end{minipage}
%
\begin{minipage}[t]{0.450\linewidth}
  \centering
  \centerline{\includegraphics[width=0.965\columnwidth]{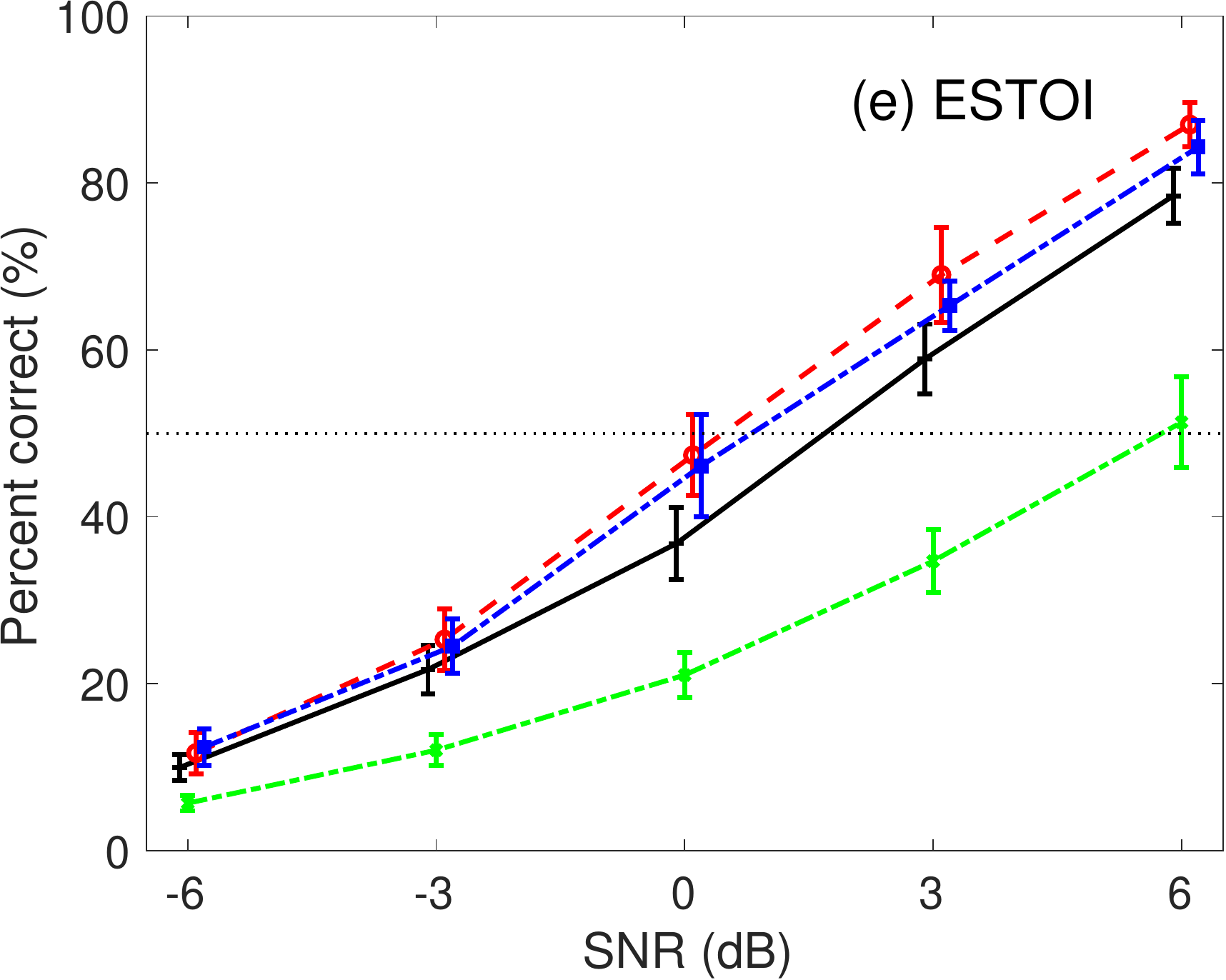}}
\end{minipage}
%
\begin{minipage}[t]{0.450\linewidth}
  \centering
  \centerline{\includegraphics[width=0.965\columnwidth]{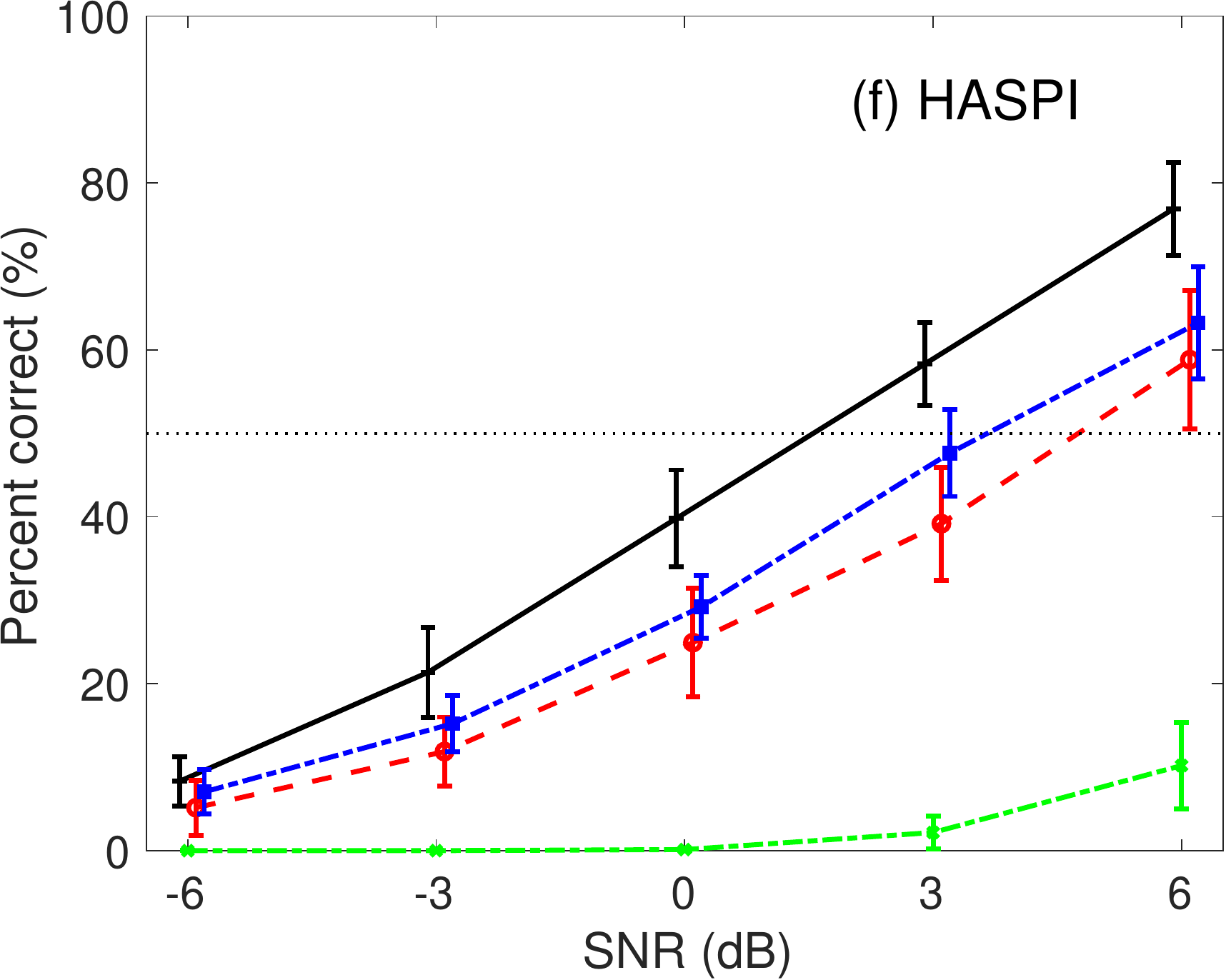}}
\end{minipage}
\caption{Results of speech intelligibility tests under babble-noise conditions. (a) Human results, the objective predictions obtained via (b) GEDI, (c) mr-GEDI, (d) STOI, (e) ESTOI, and (f) HASPI. The values and error bars represent the averages and standard deviations across the sets of speech materials.}
\label{fig:Results_PC_Babble}
\end{center}
\end{figure}

Figure\,\ref{fig:Results_PC_Babble} shows the percent correct values of speech intelligibility as a function of the speech SNR under babble-noise conditions. Panel (a) shows the human results.
The other panels indicate the results of (b) GEDI, (c) mr-GEDI, (d) STOI, (e) ESTOI, and (f) HASPI. 
The speech materials used for evaluation were unprocessed and enhanced sound samples produced by $\rm SS^{(1.0)}$ and two levels of $\mathrm{WF_{PSM}}$ (i.e., $\rm WF_{PSM}^{(0.0)}$ and $\rm WF_{PSM}^{(0.2)}$). 
Fourteen noisy speech sets were used for both subjective experiments and objective predictions. 

In the human results (Figure\,\ref{fig:Results_PC_Babble}(a)), the speech intelligibility curves for the $\mathrm{WF_{PSM}^{(0.0)}}$ and the $\mathrm{SS^{(1.0)}}$ were lower than the curve in the unprocessed speech condition. In the case of the $\mathrm{WF_{PSM}^{(0.2)}}$, speech intelligibility was similar to the unprocessed condition in every SNRs. The intelligibility score curves for the enhancement algorithms were poorly parallel to the case of SNR conditions. 
Multiple comparison analyses (Tukey--Kramer HSD test, $\alpha = 0.05$) indicated a lower speech intelligibility scores for enhanced speech processed by $\rm SS^{(1.0)} $ compared with the unprocessed speech results. There were no significant differences between the other algorithms and unprocessed speech. 

Prediction results of GEDI in  Fig\,\ref{fig:Results_PC_Babble}(b) are lower than the human results, because GEDI does not perform temporal analysis to handle non-stationary noise. 
This motivated us to develop mr-GEDI as described in Section \ref{sec:mr-GEDI}.

The prediction results of mr-GEDI (Figure\,\ref{fig:Results_PC_Babble}(c)) were improved from those of GEDI (Figure\,\ref{fig:Results_PC_Babble}(b)). However, they remain lower than the human results (Figure\,\ref{fig:Results_PC_Babble}(a)). Although improvements are necessary, the orders of the prediction curves are similar. The curve for the $\mathrm{WF_{PSM}^{(0.2)}}$ is the closest to the unprocessed curve. The $\mathrm{SS^{(1.0)}}$ curve is the most distant. The $\mathrm{WF_{PSM}^{(0.1)}}$ curve is in the achieved average results. The results may imply using mr-GEDI for further development.

In STOI (Fig\,\ref{fig:Results_PC_Babble}(d)), the prediction curves of $\mathrm{WF_{PSM}^{(0.0)}}$ and $\mathrm{WF_{PSM}^{(0.2)}}$ were higher than that of the unprocessed condition. This is inconsistent with the human results (Fig\,\ref{fig:Results_PC_Babble}(a)). The prediction curve of the $\mathrm{SS^{(1.0)}}$ was lower than that of human results.

The results of ESTOI in Fig\,\ref{fig:Results_PC_Babble}(e) were similar to STOI results. The curves for the $\mathrm{WF_{PSM}^{(0.0)}}$ and the $\mathrm{WF_{PSM}^{(0.2)}}$ were higher than the unprocessed condition curve, although the distance was smaller. 
This suggests that ESTOI predicts in a manner similar to STOI. This means that STOI can deal with non-stationary noise in the current experiments sufficiently without any extension. 

In HASPI (Fig\,\ref{fig:Results_PC_Babble}(f)), the average intelligibility of the $\mathrm{SS^{(1.0)}}$ is less than 10\%. This implies that HASPI completely failed to predict it. 
The curves for the $\mathrm{WF_{PSM}^{(0.0)}}$ and the $\mathrm{WF_{PSM}^{(0.2)}}$ were lower than the curve for the unprocessed condition. This result is similar to those of mr-GEDI.

\subsection{RMS error and bias analysis}
\label{ssec:Results_RMSE}
In the previous sections, comparisons between human and OIM results were performed qualitatively using Figures\,\ref{fig:Results_PC_Pink} and \ref{fig:Results_PC_Babble}. In this section, quantitative comparisons are performed with two measures: RMS error between the intelligibility scores of human and OIMs for individual speech enhancement algorithms and the mean difference between the unprocessed and enhanced conditions to clarify whether the prediction score is higher or lower than the unprocessed score.

\subsubsection{Pink-noise conditions}
\label{sssec:Results_RMSE_Pink}

Table\,\ref{tab:Rslt_RMSE_Pink} compares the OIMs in terms of the RMS) error between human results and predicted results for individual speech-enhancement algorithms under pink-noise conditions. Bold and italic fonts indicate the smallest and second-smallest values of each row, respectively.
The RMS error for the $\mathrm{SS^{(1.0)}}$ is the smallest of GEDI. 
The RMS errors for the $\mathrm{WF^{(0.0)}_{PSM}}$, the $\mathrm{WF^{(0.1)}_{PSM}}$, and the $\mathrm{WF^{(0.2)}_{PSM}}$ are the smallest of mr-GEDI. GEDI family successfully minimized RMS errors.

The mean differences of the speech intelligibility values between the unprocessed condition and the individual OIMs were calculated to quantify the relative locations of the curves, as shown in Figs.\,\ref{fig:Results_PC_Pink} and \ref{fig:Results_PC_Babble}. 
This is a measure used to clarify the bias between unprocessed and speech-enhancement conditions.
When the value is positive (negative), the mean score is higher (lower) than the unprocessed score. The values roughly correspond to the location of the prediction curves relative to that of the unprocessed curve. 
Table\,\ref{tab:Rslt_MeanDiff_Pink} shows the results.
In $\mathrm{SS^{(1.0)}}$, the mean difference for STOI is 
\color{black} 
the 
\color{black}
closest to that of human results, and ESTOI is the second closest.
In $\mathrm{WF^{(0.0)}_{PSM}}$, mr-GEDI is 
\color{black} 
the 
\color{black}
closest and GEDI is the second closest.
In $\mathrm{WF^{(0.1)}_{PSM}}$, GEDI is 
\color{black} 
the 
\color{black}
closest and mr-GEDI is the second closest.
In $\mathrm{WF^{(0.2)}_{PSM}}$, mr-GEDI is 
\color{black} 
the 
\color{black}
closest and ESTOI is the second closest.
It is worth noting that the mean differences of STOI, ESTOI, and HASPI were positive in the $\mathrm{WF^{(0.1)}_{PSM}}$ and the $\mathrm{WF^{(0.2)}_{PSM}}$, whereas those of the human results were negative. They overestimated the results for these two $\mathrm{WF_{PSM}}$s.

\color{black} 
In Table\,\ref{tab:Rslt_RMSE_Pink}, the RMS errors between human and predicted results displayed in Fig.\,\ref{fig:Results_PC_Pink} are shown. Smaller values correspond to better predictions from human results. The errors in GEDI and mr-GEDI were always smaller than those in STOI, ESTOI, and HASPI. However, the differences are not very large. To further clarify the difference, statistical analysis was performed as described in section\,\ref{ssec:Results_SRT}.

In Table\,\ref{tab:Rslt_MeanDiff_Pink}, the mean differences between the unprocessed and enhanced speech curves in Fig.\,\ref{fig:Results_PC_Pink} are shown. This is a bias analysis similar to that in the Bland-Altman analysis \citep{bland1986statistical}.
The results show whether the results of the enhanced algorithms were better than those of unprocessed cases. In human results in the leftmost column, the value, 2.2 in $\mathrm{WF^{(0.2)}_{PSM}}$ was positive and slightly greater than that in the unprocessed results. This implies that $\mathrm{WF^{(0.2)}_{PSM}}$ improves human performance. In contrast, the values of $\mathrm{SS^{(1.0)}}$,  $\mathrm{WF^{(0.0)}_{PSM}}$, and $\mathrm{WF^{(0.1)}_{PSM}}$ were all negative. This implies that these algorithms degrade performance.
The objective index is better when the predicted values are closer to the human results. 
In the $\mathrm{WF_{PSM}}$ family, the values of GEDI or mr-GEDI were closest to the human results. In $\mathrm{SS^{(1.0)}}$, the value of STOI is the closest. Even so, the values of GEDI or mr-GEDI were closer to the human results than the value of HASPI.

\color{black} 

\begin{table}[htbp]
\caption{RMS error  between human and predicted results in percentages under pink-noise conditions, as shown in Fig. \ref{fig:Results_PC_Pink}. Bold and italic fonts indicate the smallest and second-smallest values of each row, respectively.
}
\label{tab:Rslt_RMSE_Pink}
\begin{center}
\begin{tabular}{cccccc}
    \toprule
     & GEDI & mr-GEDI & STOI & ESTOI & HASPI \\ 
    \midrule
    $\mathrm{SS}^{(1.0)}$ &  $\textbf{10.8}$ &  $11.4$ & $\textit{10.9}$ & $11.5$ & $12.1$ \\
    $\mathrm{WF^{(0.0)}_{PSM}}$ &  $17.7$ &  $\textbf{15.1}$ & $20.0$ & $\textit{15.9}$ & $19.5$ \\
    $\mathrm{WF^{(0.1)}_{PSM}}$ &  $\textit{11.4}$ &  $\textbf{11.1}$ & $17.8$ & $14.0$ & $16.5$ \\
    $\mathrm{WF^{(0.2)}_{PSM}}$ &  $\textit{10.7}$ &  $\textbf{10.4}$ & $12.8$ & $\textit{10.7}$ & $11.1$ \\    
    \bottomrule
    \end{tabular}
\end{center}
\end{table}

\begin{table}[htbp]
\caption{
Mean difference between unprocessed and enhanced speech curves in percentages under pink-noise conditions, as shown in Fig. \ref{fig:Results_PC_Pink}. Positive (negative) values imply that the speech-enhancement algorithm improved (degraded) speech intelligibility. Bold and italic fonts indicate 
the closest and second-closest predictions
to the human results.}
\label{tab:Rslt_MeanDiff_Pink}
\begin{center}
\begin{tabular}{ccccccc}
    \toprule
     & Human & GEDI & mr-GEDI & STOI &  ESTOI & HASPI \\ 
    \midrule
    $\mathrm{SS}^{(1.0)}$ 	    &  $-13.1$ &  $-9.4$  &  $-6.7$   & $\textbf{-13.0}$    &  $\textit{-10.8}$  & $-4.8$ \\
    $\mathrm{WF^{(0.0)}_{PSM}}$ &  $-3.3$ &  $\textit{-9.9}$  &  $\textbf{-5.1}$   & $10.7$   &  $7.7$   & $12.0$ \\
    $\mathrm{WF^{(0.1)}_{PSM}}$ &  $-3.5$  &  $\textbf{-1.4}$   &  $\textit{1.3}$  & $10.7$   &  $6.4$   & $9.8$ \\
    $\mathrm{WF^{(0.2)}_{PSM}}$ &  $2.2$   &  $-3.7$   &  $\textbf{5.9}$    & $10.2$   &  $\textit{6.3}$   & $8.6$ \\ 
    \bottomrule
    \end{tabular}
\end{center}
\end{table}

\subsubsection{Babble-noise conditions}
\label{sssec:Results_RMSE_Babble}

Table\,\ref{tab:Rslt_RMSE_Babble} shows a comparison of the OIMs in terms of the RMS error under babble-noise conditions. 
\color{black} 
In $\mathrm{SS^{(1.0)}}$, STOI and ESTOI possess the smallest RMS errors. 
\color{black}
In $\mathrm{WF^{(0.0)}_{PSM}}$, the RMS error of HASPI is 
\color{black} 
the 
\color{black}
smallest, and that of mr-GEDI is 
\color{black} 
the 
\color{black}
second smallest. In $\mathrm{WF^{(0.2)}_{PSM}}$, the RMS error of ESTOI is 
\color{black} 
the 
\color{black}
smallest, and that of STOI was 
\color{black} 
the 
\color{black}
second smallest.
\color{black}

Table\,\ref{tab:Rslt_MeanDiff_Babble} shows the mean difference under babble-noise conditions. 
\color{black} 
In the $\mathrm{SS^{(1.0)}}$, STOI was 
\color{black} 
the 
\color{black}
closest to the human result, and ESTOI was 
\color{black} 
the 
\color{black}
second closest. In $\mathrm{WF^{(0.0)}_{PSM}}$, HASPI was 
\color{black} 
the 
\color{black}
closest, and mr-GEDI was the second closest. In $\mathrm{WF^{(0.2)}_{PSM}}$, ESTOI was 
\color{black} 
the 
\color{black}
closest, and mr-GEDI was the second closest. 

\color{black} 
One problem is observed from the results of STOI and ESTOI, as shown in Table\,\ref{tab:Rslt_MeanDiff_Babble}. In the mean differences of STOI, 
\color{black}
the value, $9.9$ in $\mathrm{WF^{(0.0)}_{PSM}}$, is positive and greater than the value, $7.7$ in $\mathrm{WF^{(0.2)}_{PSM}}$. This is inconsistent with human results, in which the value, $-4.9$ in $\mathrm{WF^{(0.0)}_{PSM}}$, was negative and smaller than the value, $0.3$, in the $\mathrm{WF^{(0.2)}_{PSM}}$. For human listeners, speech sounds processed by the $\mathrm{WF^{(0.2)}_{PSM}}$ were easier than those of $\mathrm{WF^{(0.0)}_{PSM}}$. STOI and ESTOI predicted this oppositely, whereas it was not the case for GEDI, mr-GESI, and HASPI.

The results from this section imply that mr-GEDI, STOI, and ESTOI were competitive in predicting under babble conditions when evaluated in terms of RMS errors and mean differences.
It is also clear that mr-GEDI was properly extended from GEDI.

\begin{table}[htbp]
\caption{RMS errors between human results and predicted results in percentages under babble-noise conditions, as shown in Fig. \ref{fig:Results_PC_Babble}. Bold and italic fonts indicate the smallest and second-smallest values in each row, respectively.}
\label{tab:Rslt_RMSE_Babble}
\begin{center}
\begin{tabular}{cccccc}
    \toprule
     & GEDI & mr-GEDI & STOI & ESTOI & HASPI \\ 
    \midrule
    $\mathrm{SS}^{(1.0)}$ &  $22.3$ &  $17.0$ & $\textbf{13.4}$ &  $\textbf{13.4}$ & $34.7$ \\
    $\mathrm{WF^{(0.0)}_{PSM}}$ &  $24.4$ &  $\textit{16.8}$ & $18.9$ &  $17.1$ & $\textbf{15.7}$ \\
    $\mathrm{WF^{(0.2)}_{PSM}}$ &  $19.1$ &  $14.2$ & $\textit{12.0}$ &  $\textbf{11.3}$ & $13.9$  \\
    \bottomrule
    \end{tabular}
\end{center}
\end{table}

\begin{table}[htbp]
\caption{Mean difference between unprocessed and enhanced speech curves in percentages under babble-noise conditions, as shown in Fig. \ref{fig:Results_PC_Babble}. Bold and italic fonts indicate the closest and second-closest predictions to the human results.}
\label{tab:Rslt_MeanDiff_Babble}
\begin{center}
\begin{tabular}{ccccccc}
    \toprule
     & Human & GEDI & mr-GEDI & STOI & ESTOI & HASPI \\ 
    \midrule
      $\mathrm{SS}^{(1.0)}$     & $-12.3$ & $-27.8$ & $-21.5$   & $\textbf{-15.3}$  & $\textit{-16.2}$  & $-38.4$  \\
    $\mathrm{WF^{(0.0)}_{PSM}}$ & $-4.9$ & $-23.1$ & $\textit{-13.6}$   & $9.9$    & $6.9$    & $\textbf{-12.9}$  \\
    $\mathrm{WF^{(0.2)}_{PSM}}$ & $0.3$  & $-13.3$ & $\textit{-6.0}$    & $7.7$    & $\textbf{5.4}$    & $-8.5$   \\ 
    \bottomrule
    \end{tabular}
\end{center}
\end{table}

\subsection{Speech reception thresholds (SRT)}
\label{ssec:Results_SRT}
Speech reception thresholds (SRT) were calculated to analyze the difference between the human and predicted results. The SRT is defined as an SNR value where the intelligibility curve crosses the 50\%-score line (dotted horizontal line in Figures\ \ref{fig:Results_PC_Pink} and \ref{fig:Results_PC_Babble}. The values of the SRTs were calculated after fitting the prediction scores with a cumulative Gaussian function. 

Moreover, $\Delta\mathrm{SRT}$ was defined to clarify the difference between the human results and the predictions by the OIMs. 
For example, $\mathrm{\Delta{SRT}}$ for the unprocessed condition is defined as
\begin{equation}
	\Delta\mathrm{SRT_{unpr.,OIM}} = \mathrm{SRT_{unpr.,OIM}} -\mathrm{SRT_{unpr.,Human}}. 
    \label{eq:dSRT}
\end{equation}
In particular, $\Delta\mathrm{SRT}$ is positive when the curve for the prediction of the OIM is located on the right side of the curve for the human result in Figures \ref{fig:Results_PC_Pink} and \ref{fig:Results_PC_Babble}. The positive $\Delta\mathrm{SRT}$ means the prediction was an under-estimation, and the negative $\Delta\mathrm{SRT}$ means the prediction was an over-estimation.

\subsubsection{Pink-noise conditions}
\label{sssec:Results_SRT_Pink}

\begin{figure}[htbp]
	\begin{center}
    \includegraphics[width=0.85\columnwidth]{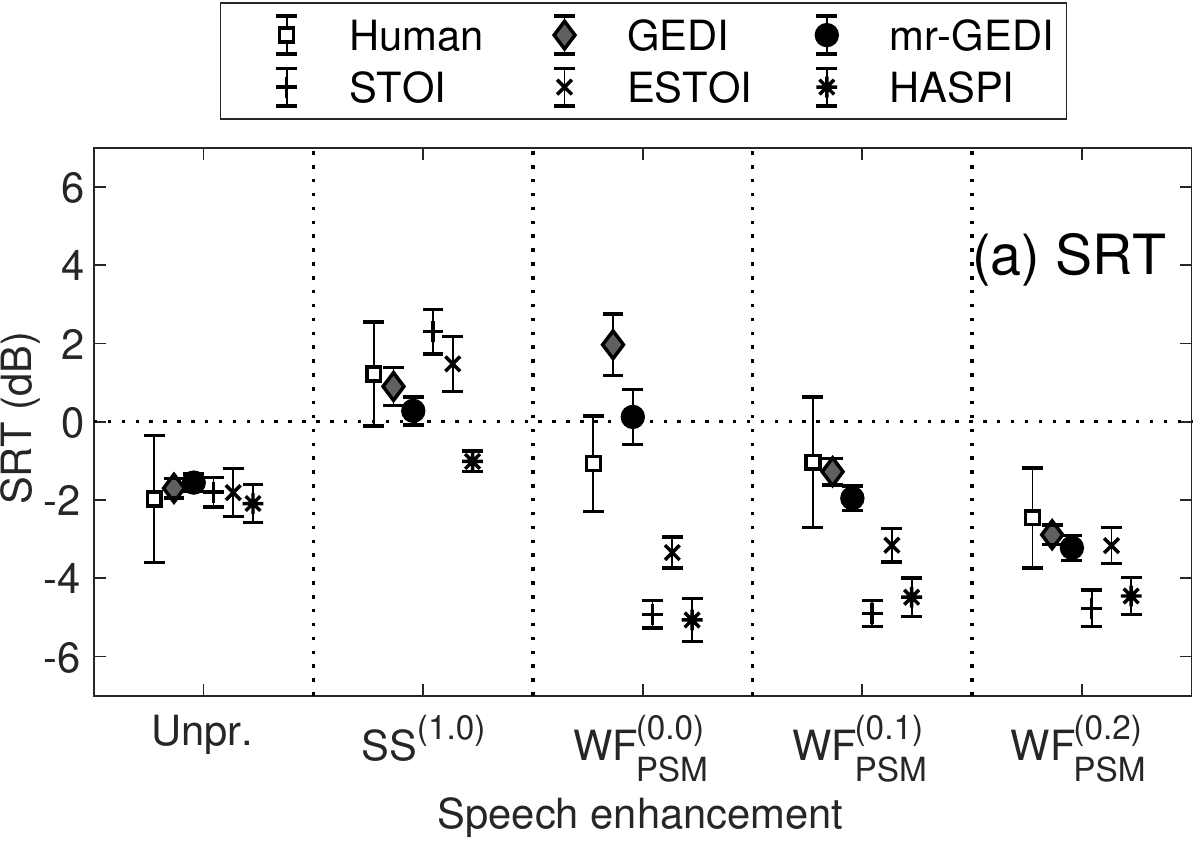}
    \includegraphics[width=0.85\columnwidth]{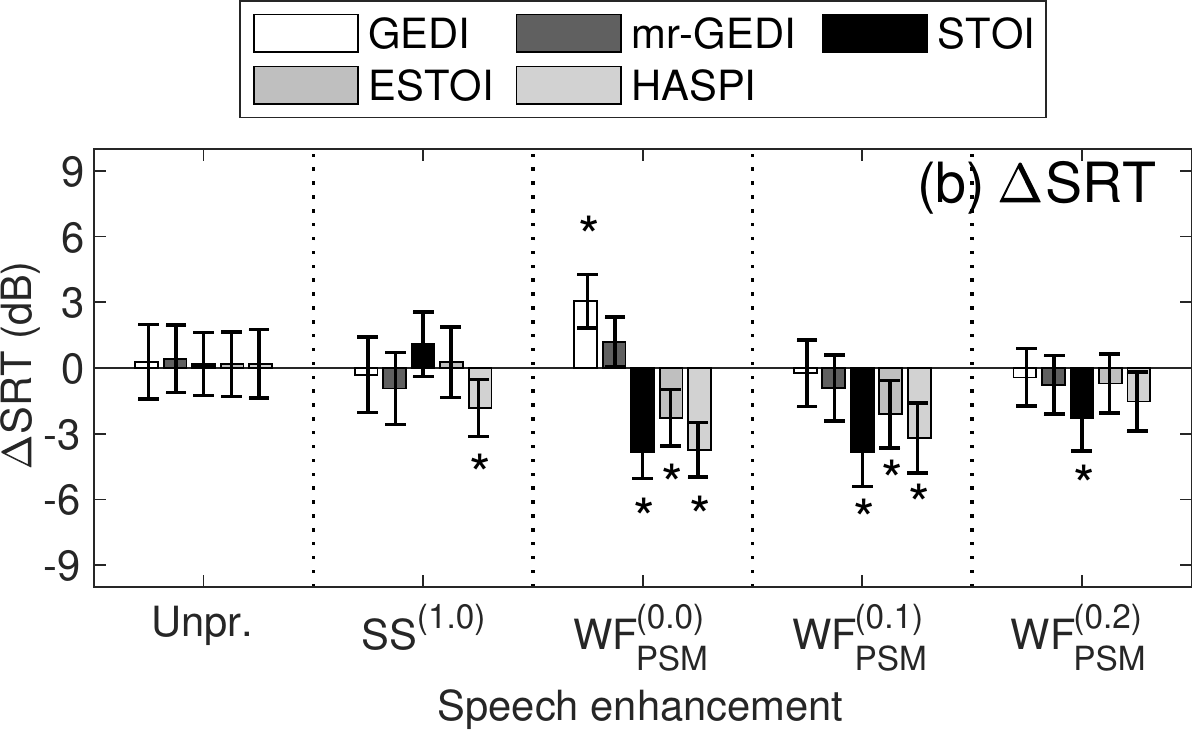}
	\caption{(a) SRT under pink-noise conditions. Marker and error bars show the mean and the standard deviation across subjects and across predictions. (b) $\Delta$SRT calculated from the SRT. Asterisk ($*$) indicates that there is significant difference (Tukey--Kramer HSD test, $\alpha = 0.05$) between the $\mathrm{\Delta{SRT}}$s for speech-enhancement and unprocessed conditions.} 
	\label{fig:Results_SRT_Pink}
	\end{center}
\end{figure}

Figure\,\ref{fig:Results_SRT_Pink}(a) summarizes the $\mathrm{SRT}$s for human results ($\square$), GEDI ($\lozenge$), mr-GEDI ($\bigcirc$), STOI ($+$), ESTOI ($\times$), and HASPI ($*$) under pink-noise conditions. 
Markers and error bars represent the mean and the standard deviations across subjects. 
In $\mathrm{SS}^{(1.0)}$, the average SRTs of mr-GEDI, GEDI, STOI, and ESTOI were within the standard deviation of the human SRT. In $\mathrm{WF_{PSM}^{(0.0)}}$, the SRT of mr-GEDI was closest to the human SRT. The SRTs of GEDI were much greater than the human SRTs, whereas the SRTs of STOI, ESTOI, and HASPI were much smaller than the human SRT. 
In $\mathrm{WF_{PSM}^{(0.1)}}$, the SRTs of GEDI and mr-GEDI were within the standard deviation of the human SRT. The SRTs of STOI, ESTOI, and HASPI were again smaller than the human SRT.
With $\mathrm{WF_{PSM}^{(0.2)}}$, the average SRTs of GEDI, mr-GEDI, and ESTOI were within the standard deviation of the human SRT, whereas those of STOI and HASPI were smaller.

Figure\,\ref{fig:Results_SRT_Pink}(b) shows $\mathrm{\Delta{SRT}}$s calculated by Eq.\,\ref{eq:dSRT} to clarify the difference between the SRTs of the OIM predictions and the human results shown in Figure\,\ref{fig:Results_SRT_Pink}(a). 
The mean, $\mathrm{\Delta{SRT}}$, in the unprocessed condition were virtually zero,  
because the parameters of the OIMs were optimized to simulate the results of listening experiments. 
Positive (negative) $\mathrm{\Delta{SRT}}$ means the OIM tends to underestimate (overestimate) speech intelligibility relative to the human results.

Multiple-comparison analysis (Tukey--Kramer HSD test, $\alpha = 0.05$) was performed between $\mathrm{\Delta{SRT}}$ of the unprocessed and enhanced conditions in each OIM. 
The asterisks in Figure\,\ref{fig:Results_SRT_Pink}(b) show different conditions from the corresponding unprocessed conditions i.e., virtually zero mean.
The $\mathrm{\Delta{SRT}}$ for GEDI was demonstrated positive results in $\mathrm{WF_{PSM}^{(0.0)}}$, indicating that GEDI has an underestimation tendency for speech intelligibility.
$\mathrm{\Delta{SRT}}$ for STOI, ESTOI and HASPI in $\mathrm{WF_{PSM}^{(0.0)}}$ and $\mathrm{WF_{PSM}^{(0.1)}}$ were significantly negative, which means that these OIMs tend to overestimate speech intelligibility. 
In contrast, $\mathrm{\Delta{SRT}}$s for mr-GEDI in every speech-enhancement conditions
were not significantly different from $\mathrm{\Delta{SRT}}$ of the unprocessed condition.
The results of the $\mathrm{SRT}$ and the $\mathrm{\Delta{SRT}}$ imply that mr-GEDI is better than STOI, ESTOI, and HASPI under pink-noise conditions. 
\color{black}
The effect of the number of listeners (sample size $= 9$) on these results is discussed in \ref{sec:NumberOfListeners}.
\color{black}

\begin{figure}[htbp]
	\begin{center}
    \includegraphics[width=0.875\columnwidth]{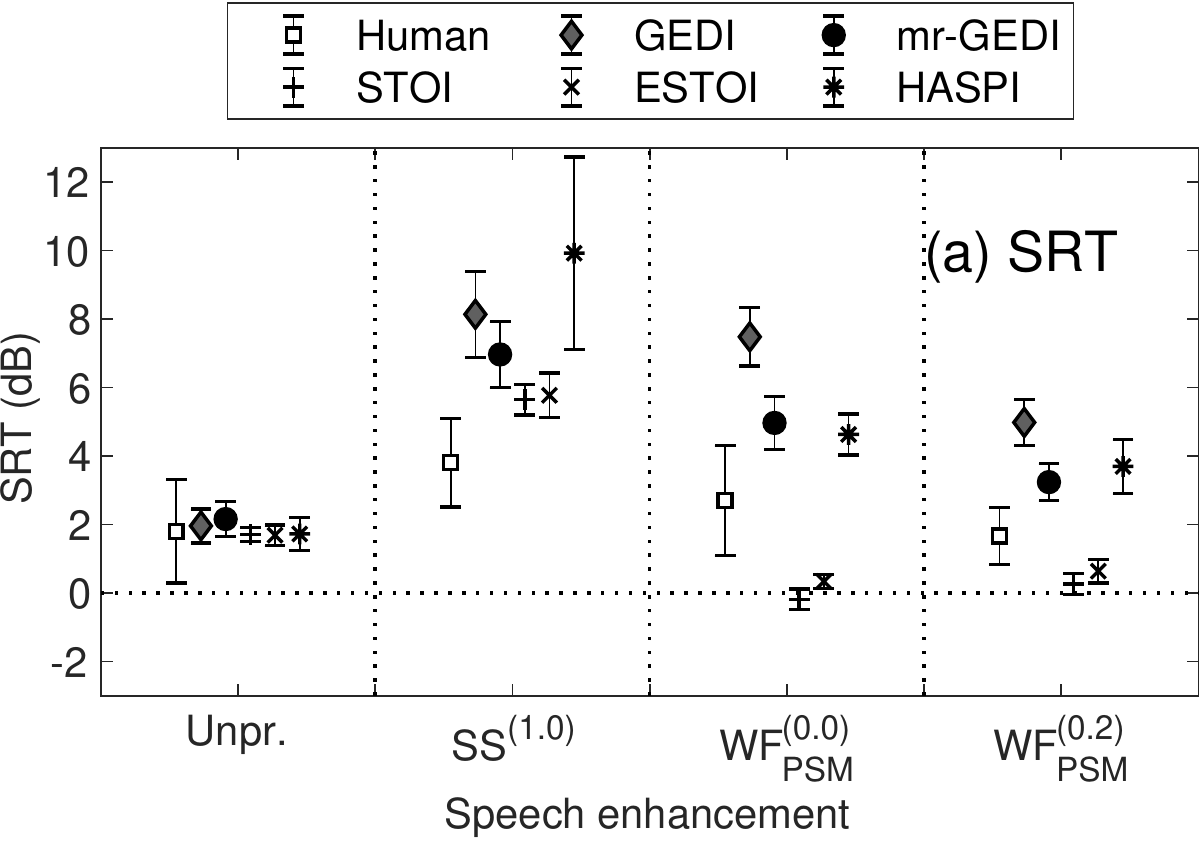}
    \includegraphics[width=0.85\columnwidth]{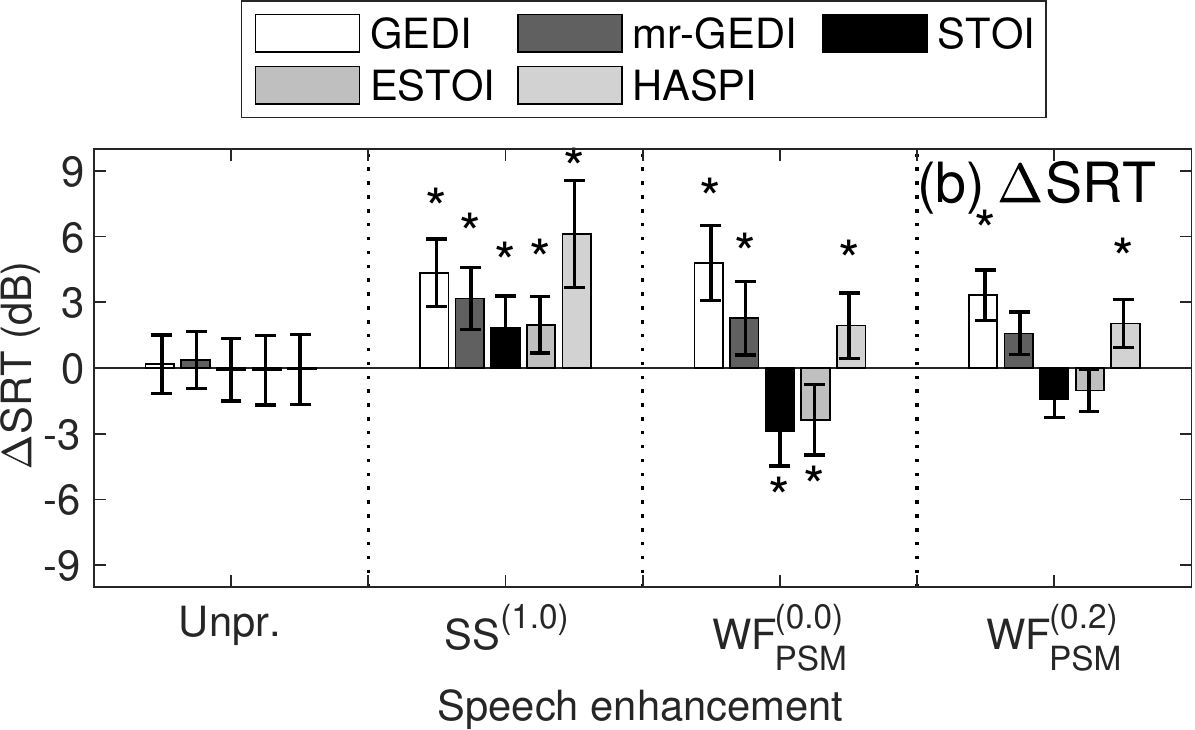}
	\caption{(a) SRT under babble-noise conditions. Markers and error bars show the mean and the standard deviation across the subjects or across predictions. (b) $\Delta$SRT calculated from SRT. Asterisk ($*$) indicates that there is significant difference (Tukey--Kramer HSD test, $\alpha = 0.05$) between the $\mathrm{\Delta{SRT}}$s for speech-enhancement and unprocessed conditions.} 
	\label{fig:Results_SRT_Babble}
	\end{center}
\end{figure}

\subsubsection{Babble-noise conditions}
\label{sssec:Results_SRT_Babble}
\color{black}
Figure\,\ref{fig:Results_SRT_Babble}(a) shows the $\mathrm{SRT}$s under babble-noise conditions. 
The $\mathrm{SRT}$s for every OIM are located inside the standard deviation of the human $\mathrm{SRT}$ in the unprocessed condition because of parameter fitting, as described previously. In contrast, the $\mathrm{SRT}$s for all OIMs are located outside of the standard deviations of the human $\mathrm{SRT}$ in every speech-enhancement condition. Thus, prediction by any of the OIMs is not highly successful. 

Figure\,\ref{fig:Results_SRT_Babble}(b) shows the $\mathrm{\Delta{SRT}}$s. 
Multiple-comparison analysis (Tukey--Kramer HSD test, $\alpha = 0.05$) was performed between $\mathrm{\Delta{SRT}}$ of the unprocessed and enhanced conditions within each OIM.  
The asterisks in Figure\,\ref{fig:Results_SRT_Babble}(b) show the conditions having significant difference from the unprocessed condition. 
There were only three conditions with no significant difference:  mr-GEDI, STOI and ESTOI in $\mathrm{WF_{PSM}^{(0.2)}}$. 
These results imply that mr-GEDI, STOI, and, ESTOI are competitive in intelligibility predictions under babble-noise conditions. But note that the $\mathrm{\Delta{SRT}}$s of STOI and ESTOI are negative for $\mathrm{WF_{PSM}^{(0.0)}}$ and $\mathrm{WF_{PSM}^{(0.2)}}$. This implies there is overestimation tendency. In contrast, $\mathrm{\Delta{SRT}}$s of other OIMs are positive, which suggests an underestimation tendency.

\subsubsection{Summary of SRT results}

Under pink-noise conditions, mr-GEDI predicted the human results better than the other OIMs. Under babble-noise conditions, mr-GEDI, STOI, and ESTOI were competitive. 
The mr-GEDI algorithm evaluates speech intelligibility for $\mathrm{WF_{PSM}}$ more conservatively than STOI and ESTOI which overestimate as indicated by negative $\mathrm{\Delta{SRT}}$s in Figs. \ref{fig:Results_SRT_Pink} and \ref{fig:Results_SRT_Babble}.
Some speech enhancement algorithms (e.g., \cite{Wang2014}) have been proposed with the main evaluation of STOI and without subjective listening tests. However, precaution should be taken due to the overestimation fragility of STOI. The evaluation with mr-GEDI may provide additional information in the development. Furthermore, it is necessary to test mr-GEDI in a wider range of algorithms and noise conditions for a more generalized conclusion and is a direction of future works.

\section{Discussion}
\label{sec:Discussion}

\color{black} 
\subsection{Goodness of the model}
\label{ssec:Discussion_GoodnessOfModel}
\color{black} 

It is essential to determine the parameter values of the OIMs in advance to predict the intelligibility of enhanced speech, as described in Section \ref{ssec:Eval_Obj}. In this study, the parameter values were derived via the LSE method to minimize the prediction error for the unprocessed conditions.

In modelling studies, goodness of model could also be measured with two factors: the number of parameters or the degree of freedom and the predictability or stability of parameter values across various conditions. The increment of parameters improves applicability to various types of data and goodness of fit to existing data. However, it does not necessarily result in improvement to performance on unknown data. A model with stable parameters is more useful in practical situations, which are more complex than laboratory conditions. 

GEDI family and STOI family required two parameters, as defined in Eqs. \ref{Sec-02_eq:GEDI_idealobserver} and \ref{Sec-04-eq:SpIntel_STOI}, whereas HASPI required, at least, three parameters, as defined in Eq.\,\ref{eq:p_HASPI}.
The number of parameters cannot be reduced because each parameter controls a different type of feature value. If one of the parameters were reduced, the prediction would fail completely.

Table \ref{tab:ParameterValues} shows the parameter values used in the evaluation. In GEDI family, $k$ values were between 1.23 and 1.50 and not overly sensitive to noise conditions. $\sigma_S$ values were about 1.8 for pink noise and about 0.7 for babble noise. $\sigma_S$ was closely related to noise stationarity. It is an interesting topic to study on presetting a reasonable value with a minimized set of listening experiments.

Parameters $a$ and and $b$ in STOI and ESTOI were similar. Thus, for example, in STOI, $a$ was -6.44 and $b$ was 4.56 under pink-noise conditions. $a$ was -8.91 and $b$ was 5.84 under babble-noise conditions. The two parameters should be tuned to account for the different noise conditions. This differs from the characteristics of the parameters used in GEDI family. Although it would require a more sophisticated algorithm to preset the parameters, it would be advantageous to know that $a$ and $b$ were consistently negative and positive. In contrast, the three parameter values in HASPI were completely different under pink- and babble-noise conditions. Moreover, parameter $C$ flips the sign. It seems difficult to preset the parameter values, because there is no consistency.

\color{black} 
\subsection{Insights to improve GEDI and mr-GEDI}
\label{ssec:ImproveGEDImrGEDI}

It does not seem that GEDI and mr-GEDI are attractive methods based on the evaluation for babble-noise conditions as shown in Tables \ref{tab:Rslt_RMSE_Babble} and \ref{tab:Rslt_MeanDiff_Babble}. 
There are some causes which can be resolved through further modifications.
One of them may be envelope fluctuations of babble noise used in the experiments. As described in section \ref{ssec:GEDI_Evaluation_Noise}, the babble noise simulates a cocktail party situation by overlapping many speech sounds. Therefore, there remain envelope modulation components similar to those of clean speech sounds, particularly after the auditory filter analysis. This component may interfere in the estimation of $\mathrm{SDR_{env}}$. 
	
One of potential solutions would be to limit the minimum value of $\mathrm{SNR_{env}}$ as proposed in \cite{Jorgensen2011,Jorgensen2013}. This may limit the output of channels where the modulation is overestimated.
Another one is to modify the weighting function in Eq.\,\ref{eq:GEDI_SDRenv_k_Weight} and to introduce an additional weighting function in \ref{eq:GEDI_Total_SDRenv_dcGC} for optimization.
However, it may require much larger amounts of data involving subjective results because of the increased number of parameters and complexity. Further study is necessary to make these modifications.

\color{black}

\section{Conclusion}
\label{sec:Conclusion}
In this study, we proposed GEDI based on the signal-to-distortion ratio in the auditory envelope, ${\rm SDR_{env}}$. The main idea behind the proposed algorithm was to calculate the distortion between the temporal envelopes of enhanced and clean speech from the output of an auditory filterbank. Moreover, GEDI was extended with a multi-resolution analysis (mr-GEDI) to improve predictions for non-stationary noise conditions.

GEDI and mr-GEDI were evaluated against well-known OIMs: STOI, ESTOI, and HASPI. Predictability of human speech intelligibility scores were evaluated for speech sounds enhanced using a simple SS and a Wiener-filtering method. Speech sounds with additive pink and babble noises with various SNRs were used for evaluation. The prediction performance of mr-GEDI was better than those of STOI, ESTOI, and HASPI under pink-noise conditions and was better than that of HASPI under babble-noise conditions. 
\color{black}
However, the differences were not very large. The main reason is that the differences in performance between speech enhancement algorithms were relatively small even in human experiments. This is unavoidable because the purpose of this study is to develop a new OIM to evaluate such advanced speech enhancement algorithms which may merely provide small but reliable steps of improvement.
In this regard, mr-GEDI exhibited certain advantages.  The mr-GEDI method did not overestimate speech intelligibility for the relatively new $\mathrm{WF_{PSM}}$ algorithm and, thus, the prediction was more conservative than that of STOI and ESTOI.  Moreover, the parameter setting of mr-GEDI was more consistent and easier than that of HASPI. 
\color{black}
Some of speech enhancement algorithms (e.g., \cite{Wang2014}) have been proposed with STOI evaluation and without subjective listening tests. The evaluation with more conservative mr-GEDI may provide additional information in the development.

Future work includes prediction evaluations using state-of-the-art speech enhancement algorithms (e.g., DNN-based approaches) and the prediction of speech intelligibility under the SPL or hearing loss conditions. 

The software for GEDI and mr-GEDI is available online: \url{https://github.com/AMLAB-Wakayama/GEDI.git}.

\section*{Acknowledgments}
This research was partially supported by JSPS KAKENHI: Grant Numbers JP25280063, JP16H01734, JP16K12464, and 17J04227. We thank anonymous reviewers for their very helpful comments and suggestions. 

\appendix

\section{Calculation of residual-noise components}
\label{app:calc_residual_noise}
The models shown in Figure\,\ref{fig:OIM}(a),  sEPSM \citep{Jorgensen2011}, mr-sEPSM \citep{Jorgensen2013}, and dcGC-sEPSM \citep{Yamamoto2019} use the residual noise as the reference signal. During the initial speech enhancement stage, the inputs were noisy signals ($S+N$) and noises ($N$), and the outputs were enhanced signals ($\hat{S}$) and residual noises ($\tilde{N}$). During the SS method, the power of noise components was subtracted with a power estimated from a non-speech segment of noisy speech, as in Eq.\,\ref{eq:SpecSub}. The residual noise component, $P_{\tilde{N}}(f)$, can be calculated directly by subtracting the power of the estimated noise, $\tilde{P}_{N}(f)$, from the power of the original noise, $P_{N}(f)$, as in %
\begin{equation}
    P_{\tilde{N}}(f) = P_{N}(f) - \alpha \tilde{P}_{N}(f).
    \label{eq:residual_noise_ss}
\end{equation}
This provides a sufficiently simple and clear definition for the residual noise.

It is, however, difficult to define the residual noise uniquely using Wiener filtering methods. These methods use a gain function, $G(f)$, to enhance speech components. Thus, there is no direct estimation of noise components. We can assume at least two ways to define the residual noise, $P_{\tilde{N}}(f)$. The first candidate is 
\begin{equation}
    P_{\tilde{N}}(f) = G(f) \cdot P_{N}(f).
    \label{eq:residual_noise_wf1}
\end{equation}
Residual noise power is directly derived from the estimated noise power with the gain function used in speech enhancement.
The second candidate is 
\begin{equation}
    P_{\tilde{N}}(f) = P_{N}(f) - G(f) \cdot P_{S+N}(f). 
    \label{eq:residual_noise_wf2}
\end{equation}
The power of the observed noisy signal, $P_{S+N}(f)$, is processed by the gain function, $G(f)$. The residual noise power is derived by subtracting this value from the original noise power.
These definitions are not very convincing when compared to definitions in the SS. Moreover, there could be other definitions to consider.

The problem of this ambiguity motivated us to develop GEDI, which uses clean speech as the reference sound. However, models using residual noise would be useful when extending to non-intrusive OIMs while resolving the ambiguity problem. 

\section{The effect of the weighting function in GEDI}
\label{app:Effect_of_Wi}
\setcounter{figure}{0}

\begin{figure}[htbp]
\begin{center}

\begin{minipage}[t]{0.450\linewidth}
  \centering
  \centerline{\includegraphics[width=0.965\columnwidth]{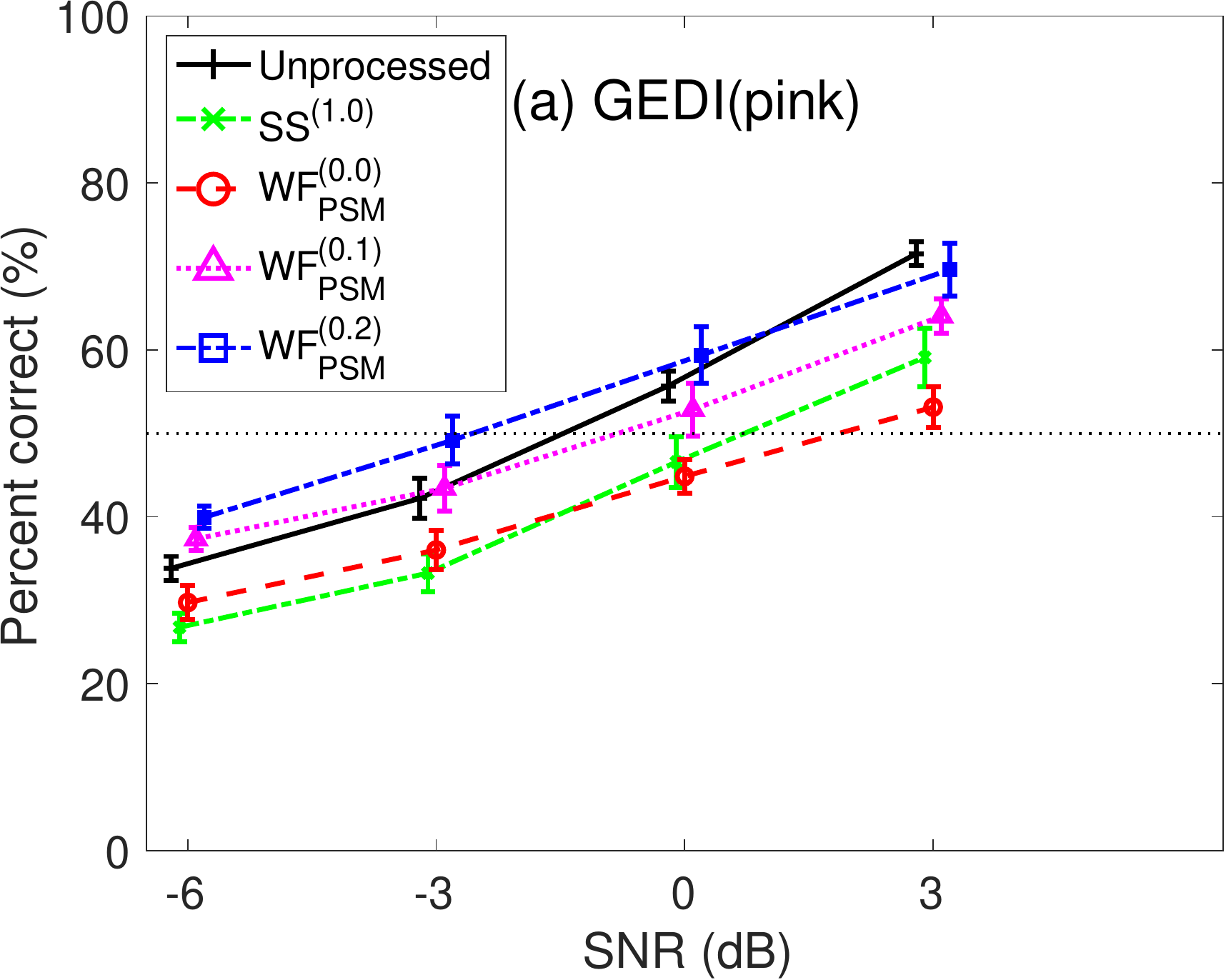}}
\end{minipage}
\begin{minipage}[t]{0.450\linewidth}
  \centering
  \centerline{\includegraphics[width=0.965\columnwidth]{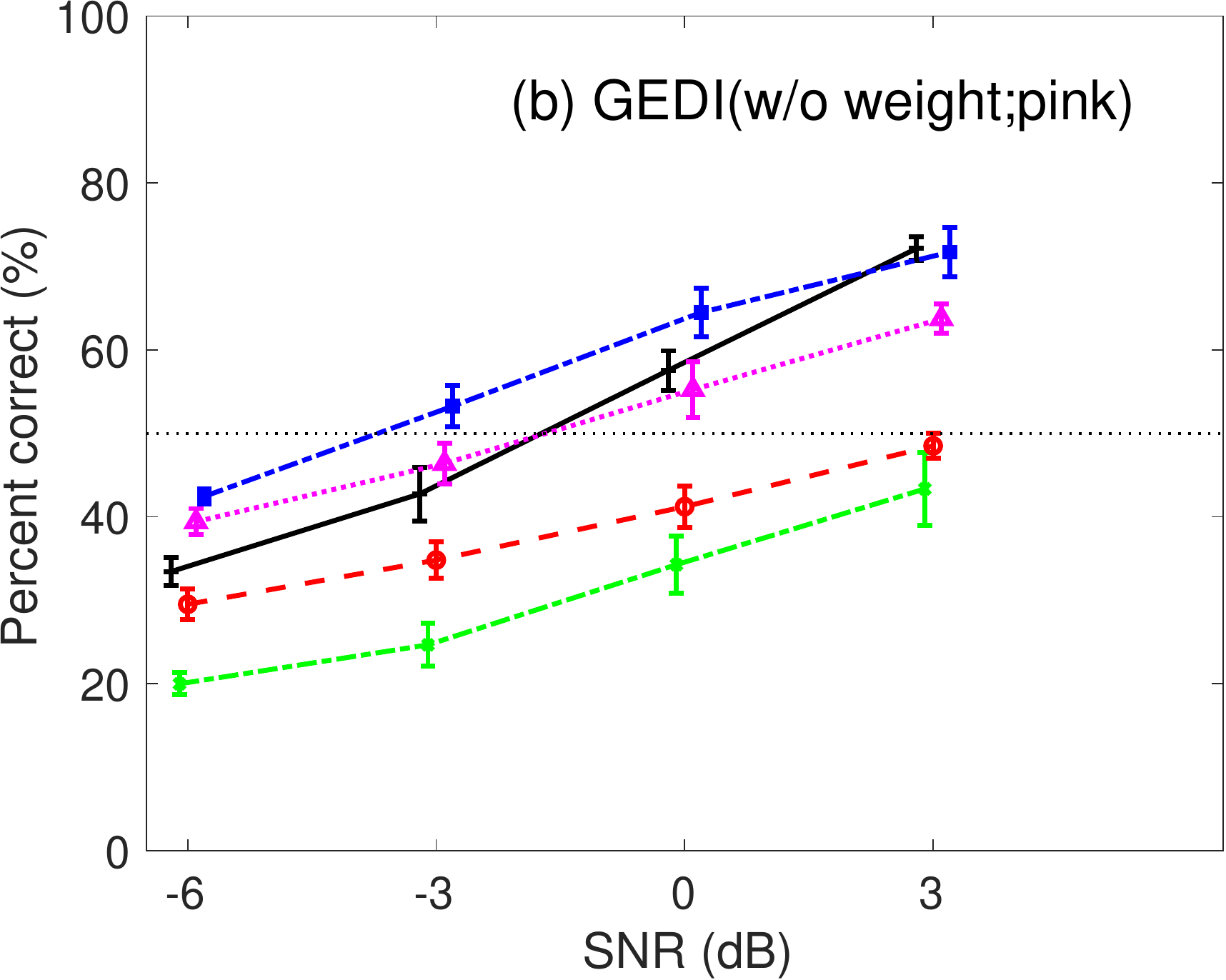}}
\end{minipage}

\begin{minipage}[t]{0.450\linewidth}
  \centering
  \centerline{\includegraphics[width=0.965\columnwidth]{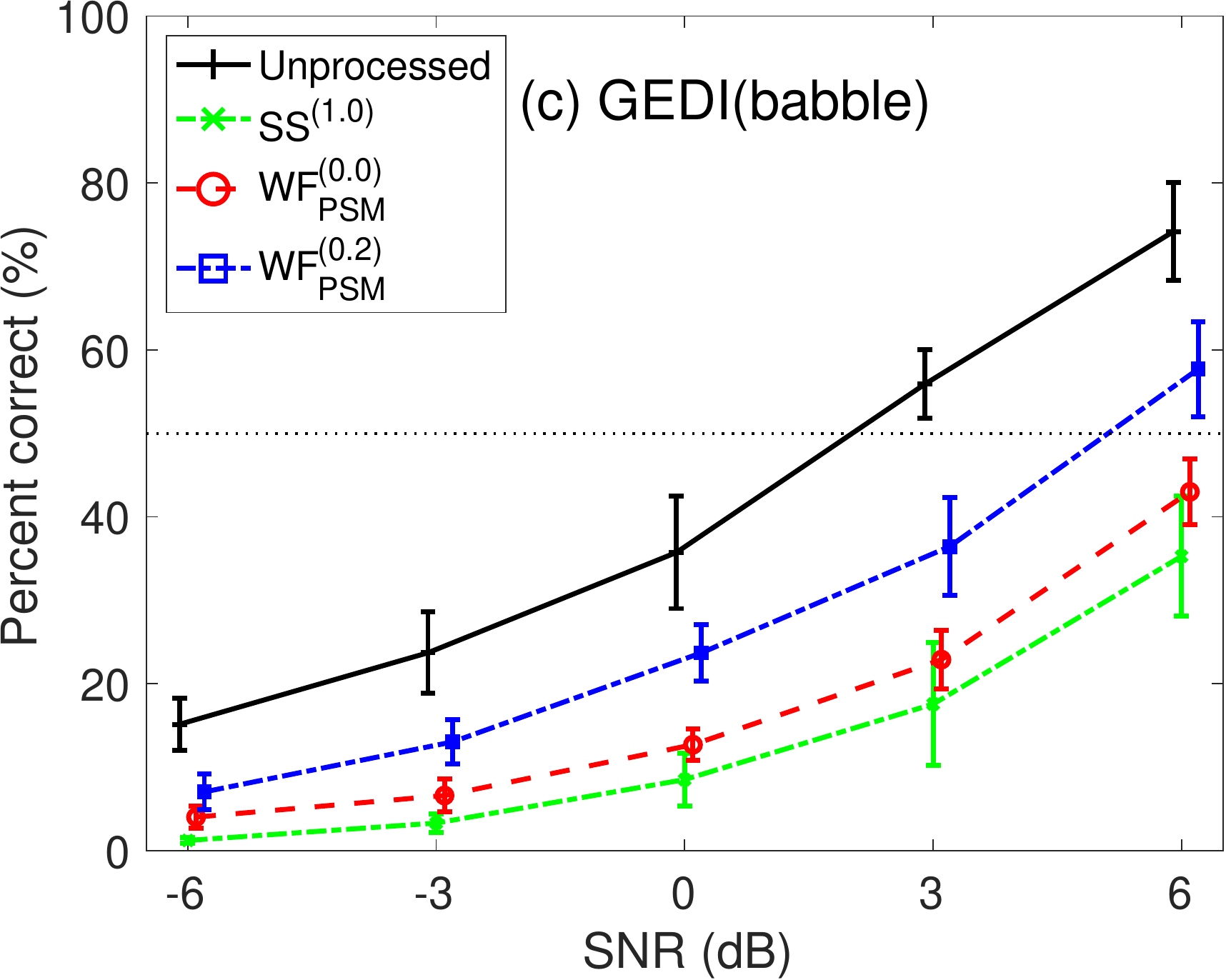}}
\end{minipage}
\begin{minipage}[t]{0.450\linewidth}
  \centering
  \centerline{\includegraphics[width=0.965\columnwidth]{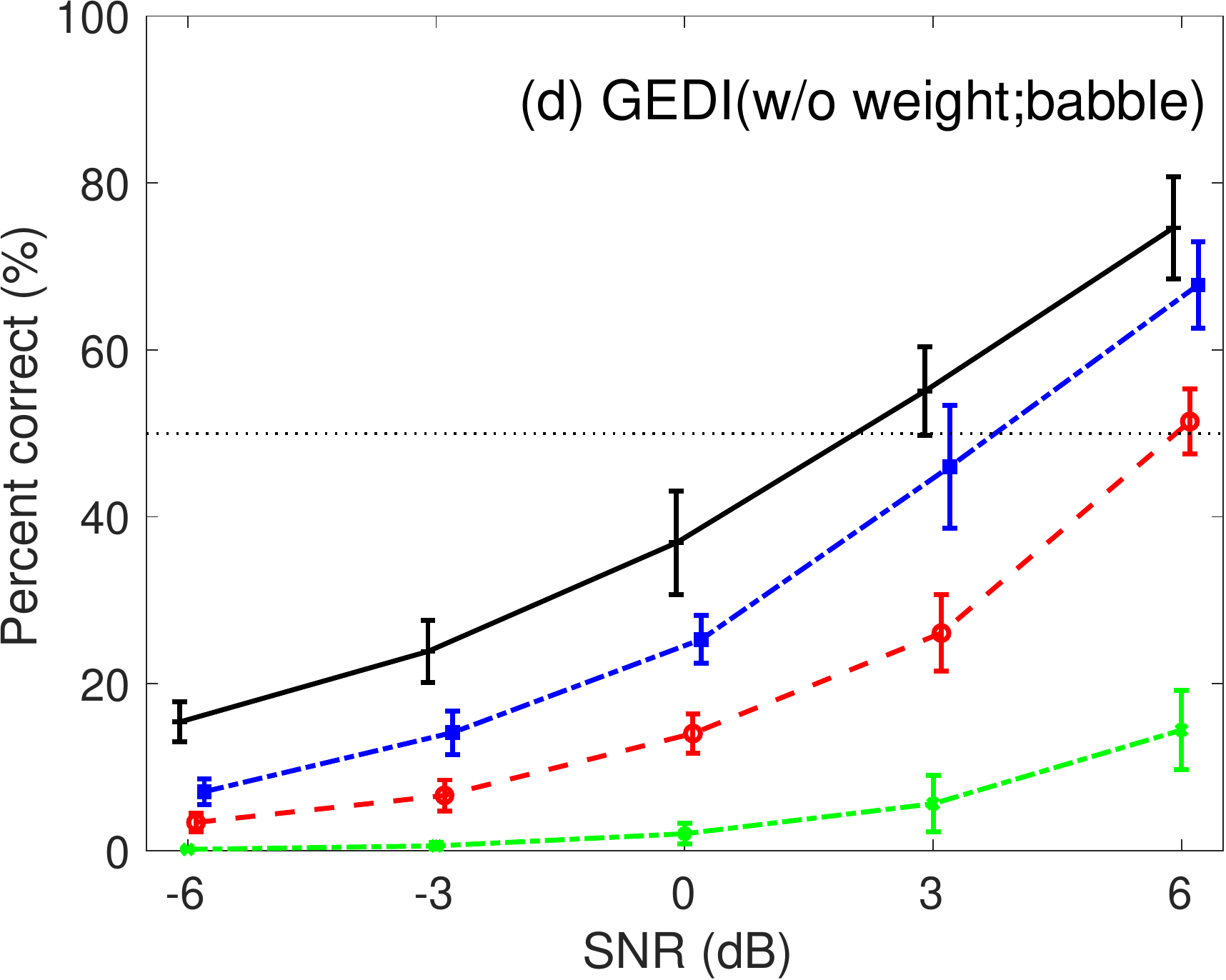}}
\end{minipage}
\caption{Prediction results of speech intelligibility tests under pink- and babble-noise conditions. (a) GEDI (with weight and pink noise: the same as Fig. \ref{fig:Results_PC_Pink}(b)), (b) GEDI (w/o weight and with pink noise), (c) GEDI (with weight and babble noise: the same as Figs. \ref{fig:Results_PC_Babble}(b)), and (d) GEDI (w/o weight and with babble noise). }
\label{fig:Effect_of_Wi}
\end{center}
\end{figure}

The effect of the weighting function, $W_i$, in Eq. \ref{eq:WeightingFunction} was evaluated for speech intelligibility prediction. Figure\,\ref{fig:Effect_of_Wi} shows the percent correct values of speech intelligibility as a function of speech SNR for pink- and babble-noise conditions. Left panels show the prediction results of GEDI (with weight), where the parameter values are listed in Table \ref{tab:ParameterValues}. Right panels show the prediction results of GEDI without the weighting function, where the parameter values were set as $k = 1.17$ and $\sigma_S = 1.62$ for pink-noise conditions and $k = 1.25$ and $\sigma_S = 0.50$ for babble-noise conditions after optimization. The prediction curves of GEDI (w/o weight) were more variable. When comparing Figures \ref{fig:Results_PC_Pink}(a) and \ref{fig:Results_PC_Babble}(a), GEDI (with weight) predicted the human results better. It is therefore clear that the weighting function, $W_i$, in Eq. \ref{eq:WeightingFunction} is important for better prediction.

\color{black}

\section{Number of listeners and its effect on the results}
\label{sec:NumberOfListeners}

The results of pink-noise conditions in Fig. \ref{fig:Results_PC_Pink}  and babble-noise conditions in Fig. \ref{fig:Results_PC_Babble} were derived from experiments with nine and fourteen listeners, respectively (see section \ref{ssec:SbjMsr_Sbj}).The number was smaller in pink-noise conditions simply because the experiment was performed earlier \citep{Yamamoto2016} than the experiment using babble noise \citep{Yamamoto2017}. As it was difficult to increase the number later, we presented the results based on that experiment.

In this section, we consider the effect of sample size on the results described in section \ref{sec:Results}. When the population mean is $\mu$, the mean and standard deviation of $n$ sample data are $\bar{x}$ and $s$, respectively, the confidence coefficient is $1-\alpha$, and the confidence interval $CI(n)$ of $\mu$ is derived (e.g. \cite{altman2013statistics}) from
\begin{eqnarray}
    CI(n)  	& = & \bar{x} \pm t_{\alpha/2}(n-1)\cdot s/\sqrt{n}\\
    		& = & \bar{x} \pm CI_{range}(n),
    \label{eq:CI}
\end{eqnarray}
where $t_{\alpha/2}(n-1)$ is the appropriate value from the $t$ distribution of $n-1$ degrees of freedom associated with $\alpha$. The ratio between the ranges of 9 and 14 is calculated as $  CI_{range}(9)/CI_{range}(14)=\sqrt{14/9}\cdot t_{0.05/2}(8)/t_{0.05/2}(13) \simeq 1.33$ when $\alpha=0.05$ and the values of $s$ are the same\footnote{This assumption about $s$ is not so unreasonable because the standard deviations of human results were roughly the same in the pink-noise and babble-noise conditions as shown in  Figs. \ref{fig:Results_PC_Pink}(a) and \ref{fig:Results_PC_Babble}(a).}.
This result means that the range of the CI is greater and the power of the test is smaller when $n=9$ than when $n=14$. In the case of the statistical analysis of $\Delta$ SRT under pink noise in Fig. \ref{fig:Results_SRT_Pink}, there were nine conditions, marked by asterisks ($*$), which were significantly different from the unprocessed conditions. If we increased the number from 9 to 14 and the range of the CI is reduced, the hypotheses of these conditions would be more strictly rejected. In contrast, the conditions where zero is contained within the range of $\bar{x} \pm s$ would not be affected much. Many of GEDI and mr-GEDI conditions exhibit these cases.  In this sense, the current analysis would not be so unreliable even though $n=9$. Even if the null hypothesis is not rejected statistically, it does not mean that the data was sampled from a population with zero-mean. But it is better than the rejected cases.

\color{black}

\section*{References}
\bibliography{SpCom19_YIAKN}

\end{document}